\documentclass[12pt,twoside,a4paper]{article}
\usepackage{amsmath,amssymb,latexsym,theorem,natbib,epsfig,color,subfigure,footmisc,bbm}
\usepackage{multirow,graphicx,array,rotating,url}  
\usepackage{epstopdf,afterpage,wasysym}
\usepackage{verbatim}
\usepackage{wrapfig}
\usepackage{natbib}
\usepackage{hyperref}
\usepackage[normalem]{ulem}
\numberwithin{equation}{section}

\title{Statistical post-processing of operational dual-resolution wind-speed ensemble forecasts}

\author{{S\'andor Baran}$^{*}$ and {M\'aria Lakatos}\vspace*{0.5cm}\\
{\small Faculty of Informatics, University of Debrecen, Hungary}
}  

\date{}

\begin{document}

\maketitle

\footnotetext[1]{Corresponding author: \url{baran.sandor@inf.unideb.hu}}
\begin{abstract}

Weather forecasting presents several challenges, including the chaotic nature of the atmosphere and the high computational demands of numerical weather prediction models. To achieve the most accurate predictions, the ideal scenario involves the lowest possible horizontal resolution and the largest ensemble size. This study provides a detailed comparative analysis of the forecast skill of the raw and post-processed medium- and extended-range wind-speed ensemble forecasts of the European Centre for Medium-Range Weather Forecasts issued at 9 km and 36 km horizontal resolutions, respectively, and their various mixtures. We utilize the ensemble model output statistic approach for forecast calibration with three different spatial training data selection techniques. First, we investigate the performance of the 50-member medium-range and 100-member extended-range predictions -- referred to as high and low resolution, respectively -- and their 150-member dual-resolution combination. Further, we examine whether the performance of raw and post-processed low-resolution forecasts can be improved by incorporating high-resolution ensemble members. Our results confirm that all post-processed forecasts outperform the raw ensemble predictions in terms of probabilistic calibration and point forecast accuracy and that post-processing considerably reduces the differences between the various configurations. We also show that spatial resolution is superior to the ensemble size; augmenting a sufficiently large ensemble of high-resolution forecasts with low-resolution predictions does not necessarily result in a gain in forecast skill. However, our study also highlights the clear benefit of the other direction, namely, incorporating high-resolution members into low-resolution ensemble forecasts, where the most significant gains are observed in configurations with the highest number of high-resolution members.

\end{abstract}

\section{Introduction}
 \label{sec1}

The advent of ensemble forecasting systems represented a major breakthrough in meteorology, fundamentally changing the way weather predictions are generated and interpreted. Unlike traditional single-run deterministic models, which produced a single outcome based on fixed initial conditions and often overlooked atmospheric uncertainties, ensemble systems introduced a transformative approach. Ensemble forecasts are the outputs of multiple runs of a numerical weather prediction model, each starting from slightly different initial conditions or parameterizations. Their purpose is to account for uncertainties in the atmospheric system and to provide a broader, probabilistic view of expected weather conditions rather than point forecasts. The European Centre for Medium-Range Weather Forecasts (ECMWF) is one of the most advanced weather centers, offering highly accurate models for various time ranges, produced with the Integrated Forecasting System \citep[IFS;][]{ecmwf24}. Their medium-range forecasts are generated four times a day, covering forecast horizons of 1–15 days, while longer-term forecasts extend up to 46 days. The accuracy of ensemble forecasts can be influenced by various factors, one of the key ones being the resolution of the weather prediction model's grid, particularly its horizontal resolution. The upgrade of ECMWF’s IFS to Cycle 48r1 brought substantial improvements to medium-range forecasts ({\it ENS}), including an increase in horizontal resolution from 18 km to 9 km ($\text{T}_{\text{CO}}1279$). For long-term forecasts ({\it ENS extended}), the resolution improved to 36 km ($\text{T}_{\text{CO}}319$), and the number of runs was increased from two days per week to a daily schedule\footnote{\href{https://www.ecmwf.int/sites/default/files/elibrary/012023/81379-newsletter-no-176-summer-2023.pdf}{ECMWF Newsletter No. 176 – Summer 2023}}. Models with finer grids, such as those used for medium-range forecasts, are better equipped to handle smaller-scale weather phenomena, such as localized precipitation or strong wind gusts. In contrast, long-term forecasts focus on larger-scale atmospheric and oceanic processes.

However, ensemble forecasts come with challenges, such as high computational costs and the difficulty of balancing resolution and ensemble size. Moreover, they can suffer from biases and underestimation of uncertainties \citep{buizza18}. A widely used technique to address the latter issues is ensemble model output statistics \citep[EMOS;][]{grgw}, a statistical method designed to improve the accuracy of ensemble forecasts by calibrating the raw output. EMOS estimates the relationship between the raw ensemble predictions and the actual outcomes. It involves fitting a parametric model to the forecast ensemble such as a normal distribution, and using it to adjust the ensemble mean and spread to better match observed reality. By doing so, it generates more reliable probabilistic forecasts that more accurately represent the uncertainty in the atmospheric system.

By integrating forecasts at distinct spatial resolutions (9 km and 36 km in this study), the dual-resolution approach capitalizes on the strengths of both high- and low-resolution ensembles. This method aims to combine the finer, smaller-scale details offered by medium-range forecasts with the broader, large-scale insights provided by extended-range forecasts, thereby improving the overall accuracy and reliability of predictions. Furthermore, due to the computational constraints of the IFS, achieving an optimal balance between different forecast combinations is crucial. \citet{lb20} combined lower- and higher-resolution ensemble members to improve medium-range weather forecasts while staying within computational constraints. Their findings indicated that dual-resolution ensembles optimize 2-m temperature predictions, while single-resolution ensembles are more effective for 850 hPa temperature forecasts. \citet{blszb19} subsequently examined whether the dual-resolution ensembles studied by \citet{lb20} remained superior to single-resolution ensembles after statistical post-processing. The results showed that statistical post-processing significantly reduced performance differences between various single- and dual-resolution configurations. In these studies, the authors combined 50 forecast members at $\text{T}_{\text{CO}}639$ (18 km), 200 members at $\text{T}_{\text{CO}}399$ (29 km), and 254 members (45 km) at $\text{T}_{\text{CO}}255$ resolution. Later, \citet{glhrblp19} evaluated the predictive performance of raw and post-processed dual-resolution precipitation accumulation forecasts, utilizing nonparametric calibration methods, while \citet{szgb23} assessed the censored shifted gamma EMOS approach for the statistical post-processing of $\text{T}_{\text{CO}}639$ -- $\text{T}_{\text{CO}}399$ dual-resolution ensemble forecasts of the same variable, which were derived from experimental extended-range predictions. The aim of this study is to compare raw and post-processed ECMWF operational dual-resolution 10-m wind-speed forecasts and investigate the impact of incorporating varying numbers of higher-resolution members into the calibration process. 

\begin{figure}[t!]
    \centering
    \includegraphics[width=.85\textwidth]{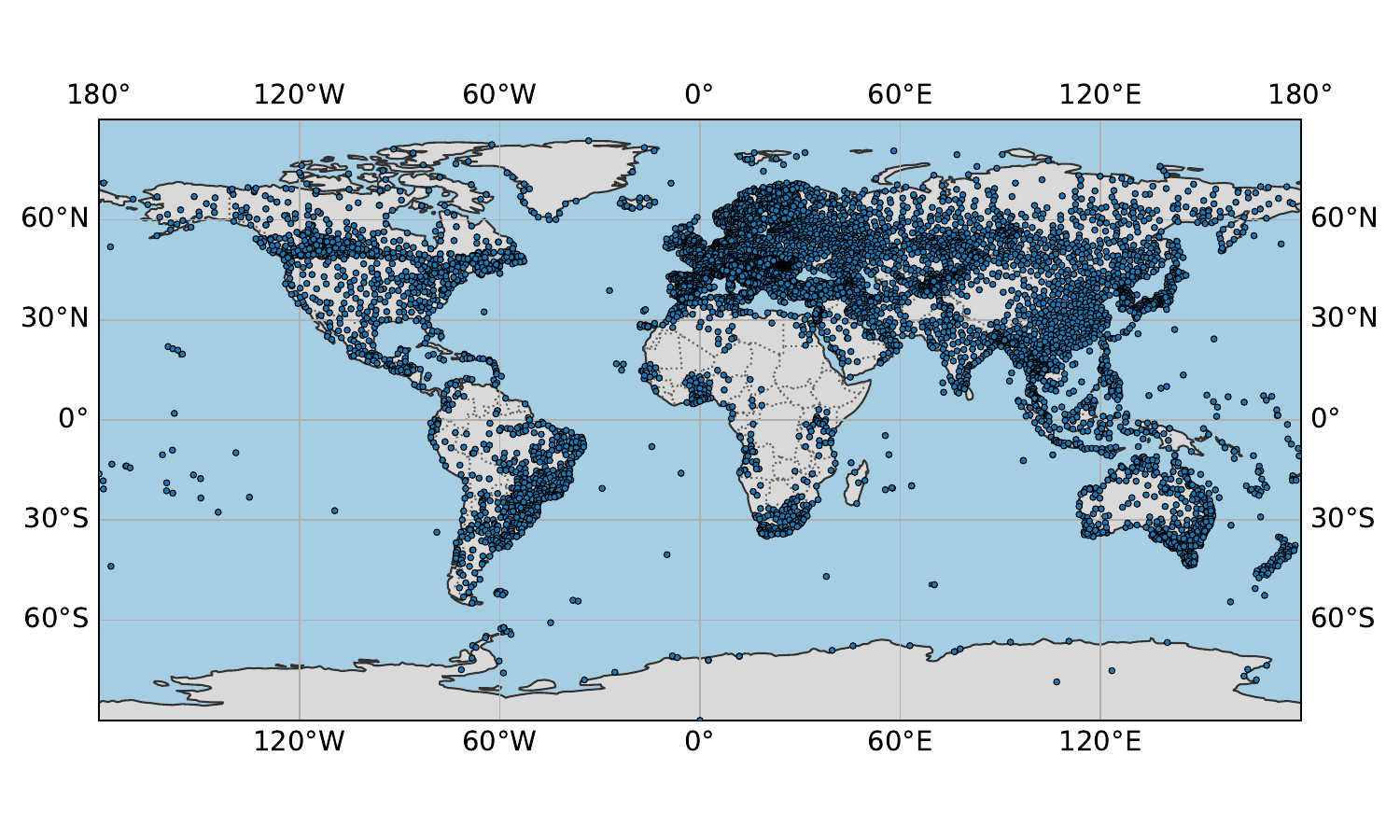}
    \caption{Location of SYNOP stations}
    \label{fig:map}
\end{figure}

First, we study the skill of raw and post-processed 50-member high-resolution and 100-member low-resolution wind-speed ensemble predictions together with their 150-member combination, where for forecast calibration, we utilize the EMOS approach. Then, we fix the number of lower-resolution members at 50, augment it with 1, 2, 4, 8, 16, and 32 higher-resolution members, and compare the predictive performance of these combinations before and after post-processing.

The structure of the paper is as follows: Section \ref{sec2} provides a detailed description of the wind-speed dataset under study. In Section \ref{sec3}, we review the calibration approaches used, along with the methods for parameter estimation and model verification. Section \ref{sec4} presents the results and Section \ref{sec5} offers the conclusions.

\section{Data}
\label{sec2}

As mentioned in the Introduction, our aim is to compare the predictive performance of ECMWF dual-resolution wind-speed forecasts and to examine whether the ranking of competing predictions would differ after calibration. 

The dataset at hand comprises 50-member medium-range forecasts of 10-m wind-speed at $\text{T}_{\text{CO}}1279$ resolution and 100-member extended-range forecasts at $\text{T}_{\text{CO}}319$ resolution, obtained from perturbed initial conditions and/or parametrizations for 8726 synoptic observation (SYNOP) stations (see Figure \ref{fig:map}) from July 1, 2023, to May 31, 2024, together with the corresponding validating observations. All forecasts are initialized at 00 UTC, and since the medium-range forecasts are limited to a 15-day horizon, the lead time of the extended-range predictions is also limited to 15 days.

\section{Statistical post-processing}
\label{sec3}

In the last two decades, a large variety of post-processing approaches have been developed both in parametric- and nonparametric setups; for a systematic overview, see \citet{vbd21}. Parametric methods provide full predictive distribution of the weather quantity at hand, whereas non-parametric approaches usually provide quantiles of the predictive law. Here, we focus on parametric post-processing, and since the aim is to explore the general tendencies in the calibration of dual-resolution forecasts, we consider the simple but still powerful EMOS approach. As mentioned, the EMOS method addresses the shortcomings of the raw predictions by applying a single parametric distribution to the ensemble outputs, where the parameters are related to the ensemble members or their descriptive statistics via appropriate (usually affine) link functions. Naturally, different meteorological variables require distinct probability distributions to best capture their unique properties. For instance, temperature is often represented by a normal distribution and its generalizations \citep{grgw, tail21}, as its values generally exhibit a symmetric spread around the mean. In contrast, wind-speed is always non-negative and skewed, making a truncated normal \citep{tg10} or a log-normal \citep{bl15} distribution a more suitable choice. Note that recently, due to its flexibility in incorporating additional input covariates, the machine learning-based counterpart of the EMOS approach, the distributional regression network \citep[DRN;][]{rl18}, has gained more and more popularity. In the DRN approach, the link functions that connect the input features to the predictive distribution parameters are replaced by a neural network so that DRN models can better capture more general relationships between these quantities, which usually results in better forecast skill. However, one should also admit that EMOS provides a well-defined statistical framework, making it more transparent and easier to interpret than machine-learning-based approaches, especially when deep neural networks are involved.

\subsection{Truncated normal EMOS model}
\label{subs3.1}

To calibrate the single- and dual-resolution wind-speed forecasts, we apply the EMOS model using a normal distribution  left-truncated at zero \ $\mathcal{N}_0^{\infty}\big(\mu, \ \sigma^2\big)$ \ with location \ $\mu$ \ and scale \ $\sigma>0$ \ \citep{tg10}. In the special case where dual-resolution ensemble forecasts are subject to calibration, the forecasts with the same horizontal resolution form distinct groups of statistically indistinguishable predictions, with \ $M_L$ \ members for the low-resolution {\it ENS extended} forecasts and \ $M_H$ \ for the high-resolution {\it ENS} predictions. Based on this, the location and scale parameters of the EMOS predictive distribution are
\begin{equation}
  \label{eq:TNemos}
\mu = a + b_H^2 \overline{f}_H + b_L^2 \overline{f}_L \qquad \text{and} \qquad \sigma^2 = c^2 + d^2 S^2,
\end{equation}
where \  $a, \ b_H, \ b_L, \ c, \ d \in {\mathbb R}$ \ are the parameters to be estimated, \ $\overline{f}_H$ \ and \ $\overline{f}_L$ \ are the means of the high- and low-resolution forecasts, respectively, while \ $S^2$ \ denotes the variance of the \ $(M_H + M_L)$-member combined ensemble. In the following sections, such a configuration of \ $M_L$ \ low- and \ $M_H$ \ high-resolution ensemble members will be referred to as a combination \ $(M_L,M_H)$. 

Alternatively, when the calibration is based solely on one resolution, the expression for location in \eqref{eq:TNemos} has to be modified accordingly by fixing \ $b_L=0$ \ for the pure high-resolution \ ($M_L=0$) \ and \ $b_H=0$ \ for the pure low-resolution \ ($M_H=0$) \ case.

Drawing on the optimal score estimation principle proposed by \citet{gr07}, the estimation of the parameters of EMOS predictive distributions involves minimizing the mean value of a proper scoring rule over a carefully chosen training dataset comprising past forecast-observation pairs. In most cases, the continuous ranked probability score (CRPS) defined by \eqref{eq:crps} in Section \ref{subs3.2} is favored, as it simultaneously evaluates the magnitude of forecast errors and the overall distributional performance, yielding a comprehensive assessment of predictive skill.

\subsection{Verification metrics}
  \label{subs3.2}

We evaluate the predictive performance of both probabilistic and deterministic forecasts using a blend of traditional error metrics and proper scoring rules. Since the post-processing methods considered are applied independently to each lead time and location, their evaluation is naturally aligned with univariate scoring rules that assess the quality of one-dimensional predictive distributions.

To assess the forecast skill of deterministic forecasts, such as the ensemble/EMOS means and medians, we utilize the root mean squared error (RMSE) and the mean absolute error (MAE), respectively. In scenarios where a single point forecast is employed to represent the future outcome, the MAE is minimized by the median of the corresponding probabilistic prediction, while the RMSE is minimized by the mean \citep{gneiting11}. 

The continuous ranked probability score \citep[CRPS; ][Section 9.5.1]{wilks19} is a proper scoring rule used to evaluate the quality of probabilistic forecasts. It measures the difference between the predicted cumulative distribution function (CDF) and the empirical CDF of the observed value. For a forecast distribution \ $F$ \ and an observation \ $y$, \ the CRPS is defined as
\begin{equation}
  \label{eq:crps}
\text{CRPS}(F, y) := \int_{-\infty}^{\infty} \left( F(z) - \mathbbm{1}\{z \geq y\} \right)^2 \, {\mathrm d}z,
\end{equation}
where \ $\mathbbm{1}\{\cdot\}$ \ denotes the indicator function. Note that for the truncated normal distribution, the CRPS has a closed form \citep[see][]{jkl19}, allowing for an efficient parameter estimation.

In addition, we include the Brier score \citep[BS; ][Section 9.4.2]{wilks19} to evaluate forecast skill for binary events derived from continuous variables. Specifically, we assess the case in which the event of interest is whether the observed value \ $y$ \ exceeds a given threshold \ $z$. \ The Brier score compares the forecast probability \ $1-F(z)$ \ of this event with its actual outcome and is defined as
\[
\text{BS}(F, y; z) := \left(F(z) - \mathbbm{1}\{z \geq y\} \right)^2.
\]
Moreover, the CRPS can be interpreted as the integral of the Brier score over all possible threshold values. In Section \ref{sec4}, following e.g. \citet{vwds21}, we consider thresholds 5 m/s, 10 m/s, and 15 m/s, corresponding to low, moderate, and high wind-speeds, respectively.

Furthermore, let \ \( q_{\tau}(F) \) \ denote the \ \(\tau\)-quantile \ \((0 \leq \tau \leq 1)\) \ of a CDF \ \(F(y)\), \ that is
\[
q_{\tau}(F) := F^{-1}(\tau) := \inf \{ y : F(y) \geq \tau \}.
\]
Consider the loss function
\[
\rho_{\tau}(y) := 
\begin{cases}
\tau |x|, & \text{if } x \geq 0, \\
(1 - \tau) |x|, & \text{if } x < 0.
\end{cases}
\]
Then, for an observed value \ \(y\), \ the quantile score \citep[QS; see e.g.][]{bf14} is defined as
\[
\text{QS}_{\tau}(F, y) := \rho_{\tau}\bigl(y - q_{\tau}(F)\bigr).
\]
In this study, we evaluate the QS at the 5th, 10th, 20th, 80th, 90th, and 95th percentiles of the predictive distributions.

The improvement of a forecast with respect to a reference predictive distribution \ \( F_{\mathrm{ref}} \) \ can be quantified using corresponding skill scores. For a generic scoring rule \ \( {\mathcal S} \), \ the skill score comparing forecasts \ \( F \) \ and \ \( F_{\mathrm{ref}} \) \ is defined as
\[
\mathcal{SS}\big(F, F_{\mathrm{ref}}\big) := 1 - \frac{\overline{{\mathcal S}_F}}{\overline{{\mathcal S}_{F_{\mathrm{ref}}}}},
\]
where \ $\overline{{\mathcal S}_F}$ \ and \ $\overline{{\mathcal S}}_{F_{\mathrm{ref}}}$ \ denote the mean score values over the verification data corresponding to \ \( F \) \ and \ \( F_{\mathrm{ref}} \), \ respectively \citep{murphy73}. Naturally, skill scores differ from the original metrics by being positively oriented, so higher values indicate better forecast quality. In Section \ref{sec4}, summarizing our results, for probabilistic forecasts, we consider the continuous ranked probability skill score (CRPSS), the quantile skill score (QSS), and the Brier skill score (BSS), whereas, for point forecasts, we investigate skill scores corresponding to the MAE of the median (MAES) and the RMSE of the mean (RMSES).

To gain insight into the uncertainty in the score values and the significance of the score differences, in Section \ref{sec4}, the reported skill scores are equipped with 95\,\% Gaussian confidence intervals. The required standard deviations are derived from 2000 block bootstrap samples calculated with the help of the stationary bootstrap scheme with random block lengths following a geometric distribution with a mean proportional to the cube root of the length of the time series of skill scores \citep{pr94}.

\subsection{Training data selection}
  \label{subs3.3}

The number of training days is a critical factor affecting the stability and adaptability of the EMOS model. A shorter training period makes the model more sensitive to short-term fluctuations, which can increase the risk of overfitting. On the other hand, a longer training period can lead to more stable parameter estimates. The choice of a spatial selection strategy also plays a vital role in estimating the parameters of the EMOS model. This estimation can be based on data from a broader region, data from individual stations, or a third approach that incorporates information from similar stations. Each of these methods -- regional, local, and semi-local -- offers distinct benefits and drawbacks in terms of forecast accuracy and stability. In the regional method \citep{tg10}, the parameters of the EMOS model are estimated using data from all available stations, hence resulting in a single set of parameters for the whole ensemble domain. By using data from all stations within a larger geographic area, it could enable more stable and reliable parameter estimation. This approach is particularly beneficial in homogeneous regions where weather conditions are similar, as patterns learned from a larger dataset can be effectively applied to local forecasts. Additionally, it is computationally more efficient, as a single model is fitted for a larger region, reducing resource demand. However, its drawback is that it is less capable of capturing location-specific characteristics. In heterogeneous regions where weather conditions vary considerably, the regional model may not always provide accurate forecasts, as individual stations may have unique climatic features that differ from the model fitted to the entire area. In contrast, local learning relies solely on the data of individual stations, allowing it to better capture the special weather characteristics of a given location. Since a separate model is fitted for each station, this approach can provide more accurate forecasts in areas where the generalization of a regional model is not suitable. This is particularly useful in mountainous or coastal regions, where topographical conditions cause significant local variations in weather parameters. While in general, local modelling overperforms the regional approach, one of its drawbacks is the limited amount of data available. For the optimal training window lengths for various weather quantities, see \citet{hemri14}. If a station has little historical data, the estimated parameters may become more uncertain, and the model might overfit the training data, reducing its generalizability. Additionally, this approach is more computationally demanding, as a separate EMOS model must be fitted for each station, increasing computational costs. By partitioning the ensemble domain into smaller, similar areas, the clustering-based semi-local approach of \citet{lb17} capitalizes on the strengths of both regional and local techniques. In each of these homogeneous areas, regional modeling is applied, allowing forecasts to be fine-tuned to local conditions while still reflecting overall regional trends. To dynamically form these areas for each verification date, we employ $k$-means clustering using feature vectors proposed by \cite{lb17} that combine climatological data with the forecast errors from the raw ensemble during the training period. Ultimately, this strategy enhances the accuracy of modeling local variations and contributes to improved overall forecast performance.

\begin{figure}[t]
\begin{center}
\epsfig{file=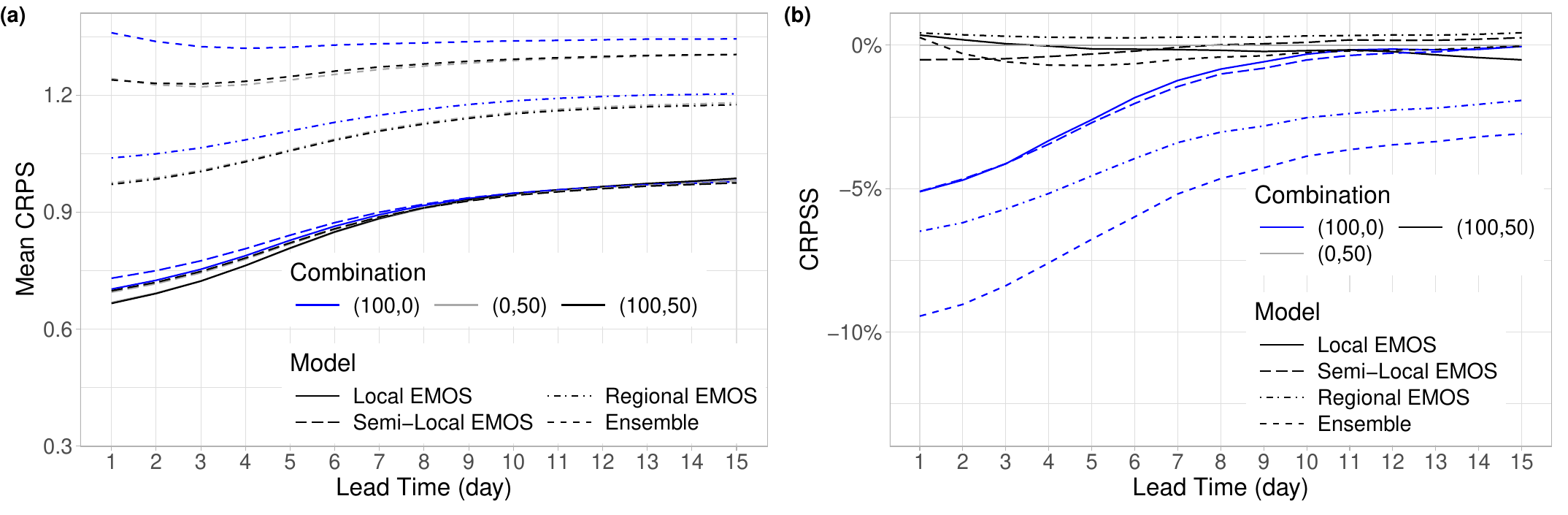, width=\textwidth}
\end{center}
\caption{Mean CRPS of pure low-resolution (100,0), pure high-resolution (0,50), and combined dual-resolution (100,50) raw and post-processed wind-speed forecasts (a) and CRPSS with respect to the pure high-resolution prediction (b) as functions of the lead time.}
\label{fig:crps_crpss_all}
\end{figure}

\section{Results}
\label{sec4}

As mentioned in the Introduction, we investigate the forecast skill of various combinations of raw high- and low-resolution wind-speed ensemble forecasts and their post-processed counterparts, where we utilize the truncated normal EMOS model introduced in Section \ref{subs3.1}. For the EMOS calibration, each lead time was modelled independently using the three distinct spatial selection strategies (regional, local, and semi-local) described in Section \ref{subs3.3}. In the semi-local approach, clustering features were extracted from the training dataset, and $k$-means clustering was performed based on these features. In particular, following the approach of \citet{lb17}, for a given location, we considered 12 equidistant quantiles of the climatological CDF and 12 equidistant quantiles of the CDF of the forecast error of the ensemble mean over the actual training period. To identify the optimal model configuration, both the number of clusters and the length of the training period were systematically tested and evaluated. Training periods of 30, 60, and 90 days were considered, and a wide range of cluster numbers was assessed within the semi-local framework. These configurations were evaluated over a dedicated verification period from 13 October 2023 to 31 May 2024, spanning 232 calendar days, to ensure comparability between different model setups. The final configuration — 90 clusters and a 60-day training period — was selected by minimizing the mean CRPS over this validation period while also considering additional verification metrics and balancing computational efficiency with predictive performance.

Based on this setup, the full verification period was defined as 3 September 2023 to 31 May 2024, totaling 262 calendar days, and was applied consistently across all model configurations during the final evaluation phase.

\begin{figure}[t]
\begin{center}
\epsfig{file=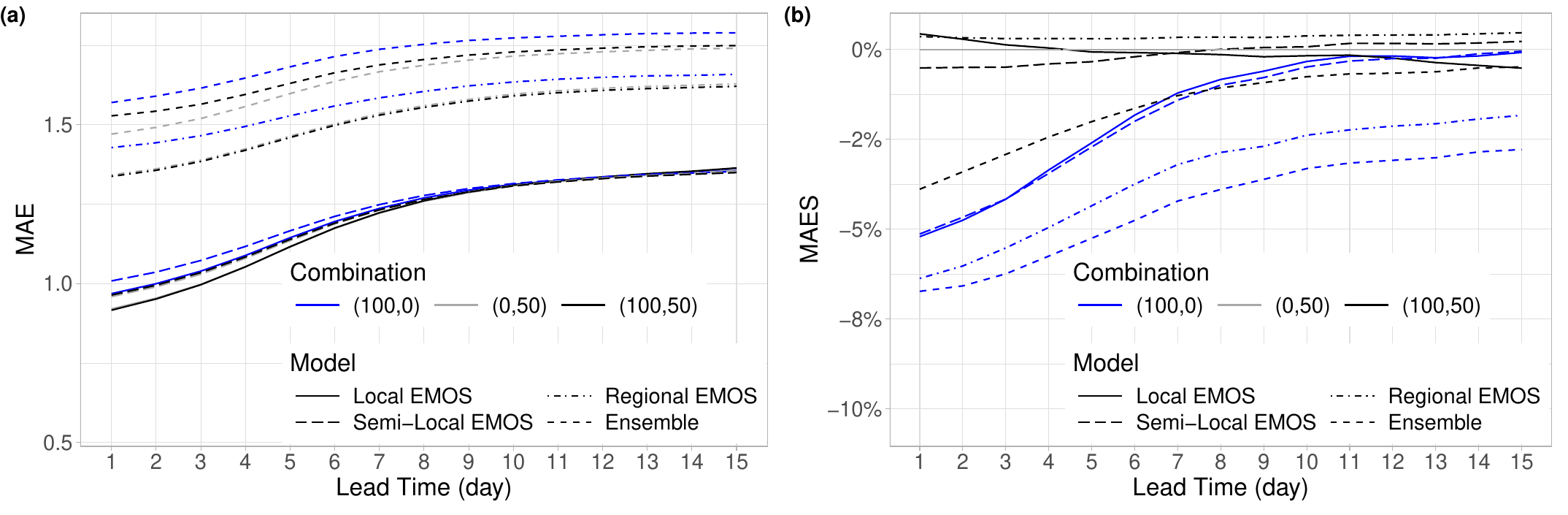, width=\textwidth}
\end{center}
\caption{MAE of the median of pure low-resolution (100,0), pure high-resolution (0,50), and combined dual-resolution (100,50) raw and post-processed wind-speed forecasts (a) and MAES with respect to the pure high-resolution prediction (b) as functions of the lead time.}
\label{fig:mae_maes_all}
\end{figure}

\subsection{Performance of operational dual-resolution forecasts}
\label{subs4.1}

In the following, we present a comparative analysis of the ECMWF operational forecasts at both high-  and low-resolution and explore their connection to dual-resolution ensemble predictions. All three configurations, referred to as combinations (0,50), (100,0), and (100,50), respectively, are post-processed using the truncated normal EMOS model, and the skill of the resulting forecasts is subsequently assessed. 

Figure \ref{fig:crps_crpss_all}a displays the mean CRPS values for raw pure low-, high-, and combined dual-resolution forecasts, alongside their post-processed counterparts, based on the spatial selection methods outlined in Section \ref{subs3.3}. Note that the non-monotonic shape of the mean CRPS curves of the raw configurations is a result of the representativeness
error in the verification \citep[see also][]{bszsz21,bl24}, which can be reduced, for instance, by perturbing the ensemble members \citep{bbhwhr20}. Among the raw configurations, the (100,0) pure low-resolution forecast exhibits the lowest predictive skill, while the (0,50) pure high- and (100,50) dual-resolution forecasts perform comparably. Figure \ref{fig:crps_crpss_all}b, where the pure high-resolution forecast is used as a reference, indicates a modest advantage of this configuration over the dual-resolution raw forecasts during the first seven to eight lead days. Among the post-processing approaches, the local and semi-local EMOS models provide the most substantial overall improvement relative to the raw forecasts. Notably, the local model outperforms during the first six forecast days, after which its performance converges with that of the semi-local approach. 

\begin{figure}[t]
\begin{center}
\epsfig{file=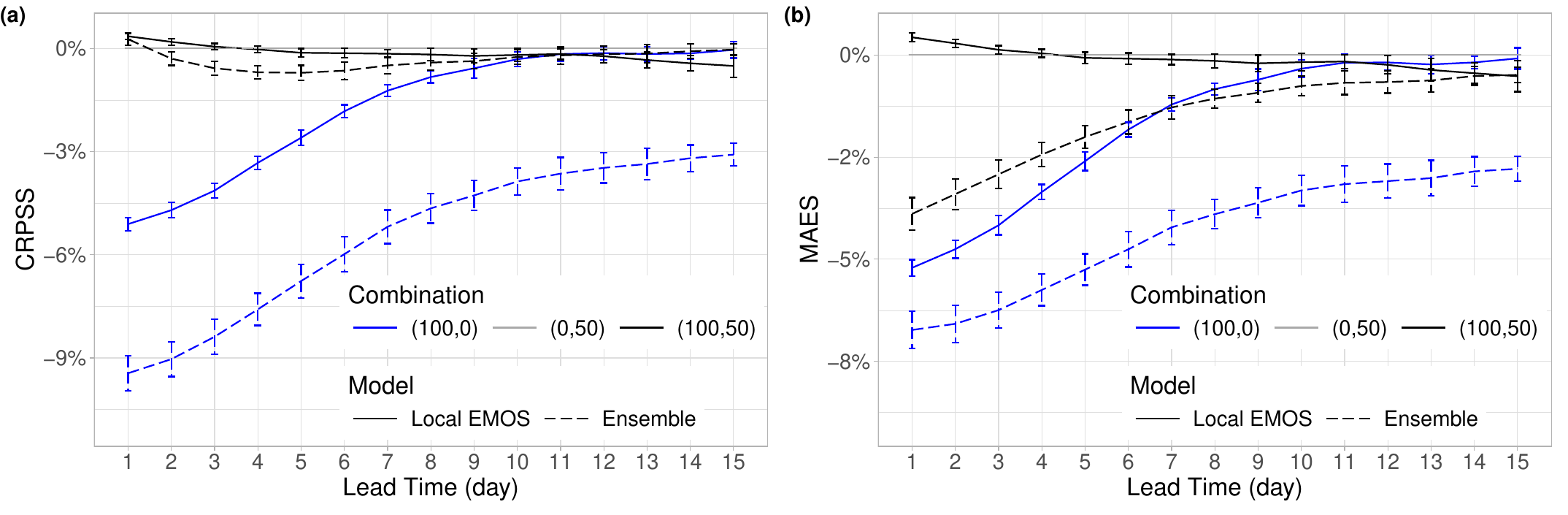, width=\textwidth}
\end{center}
\caption{CRPSS (a) and MAES (b) of pure low-resolution (100,0) and combined dual-resolution (100,50) raw and post-processed wind-speed forecasts with respect to the corresponding pure high-resolution predictions (0,50) with 95\,\% confidence intervals as functions of the lead time.}
\label{fig:crpss_maes_all}
\end{figure}

Figure \ref{fig:mae_maes_all} shows a similar overall pattern in performance differences, this time based on MAE values. Among the raw forecasts, the pure high- and dual-resolution setups again perform best, though the distinction between them is now more noticeable -- this is also reflected in panel (b). For the first few lead times, the differences between the methods are more evenly distributed. Regional post-processing offers less of an advantage for median forecasts than it did for CRPS in Figure \ref{fig:crps_crpss_all}. Meanwhile, the local and semi-local methods remain closely aligned, with the local approach consistently providing the best performance across all resolution setups, particularly during the early forecast days. Therefore, in the following analysis, we focus exclusively on this post-processing method alongside the raw forecasts. 

\begin{figure}[th!]
\begin{center}
\epsfig{file=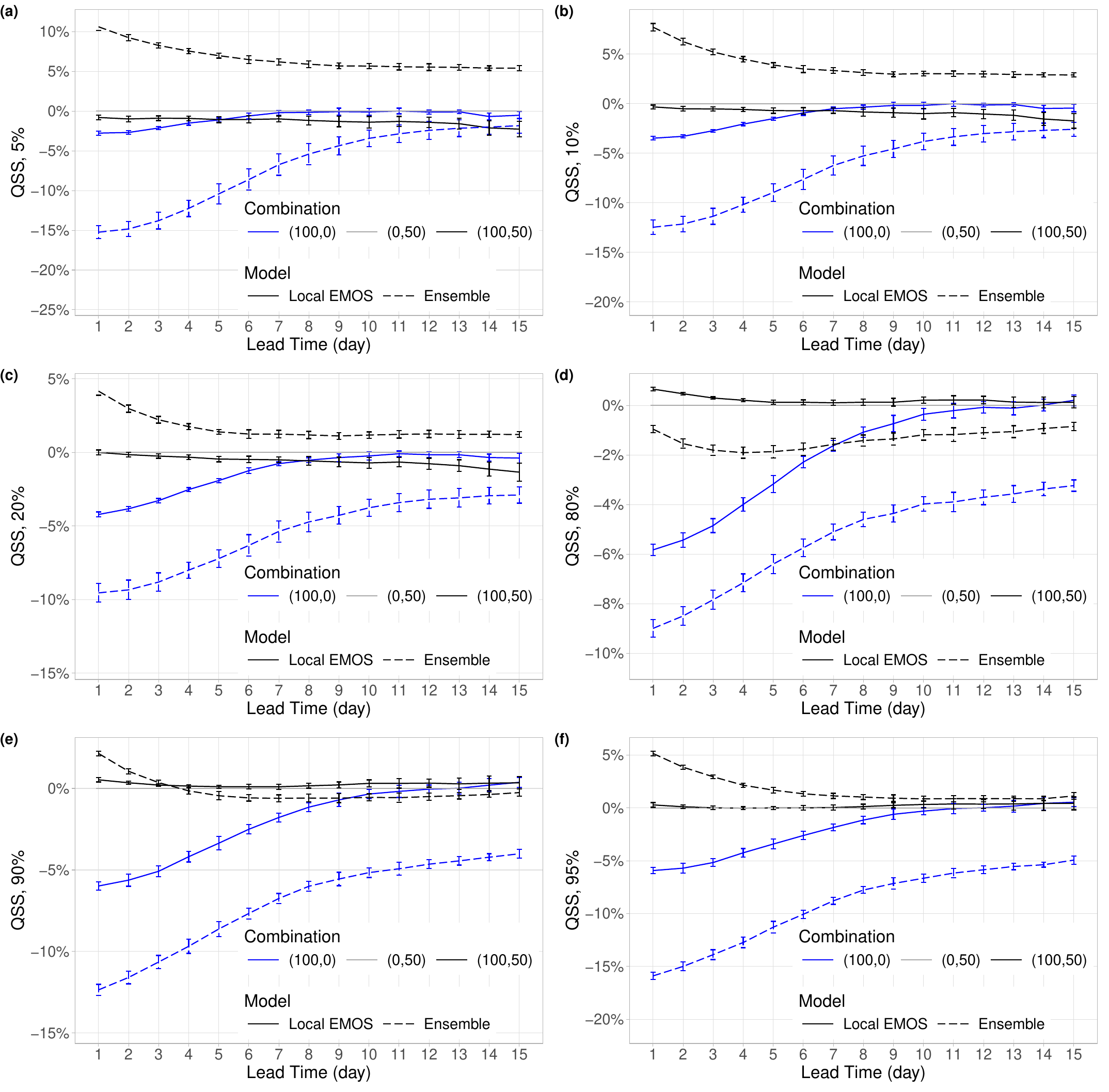, width=\textwidth}
\end{center}
\caption{QSS of pure low-resolution (100,0) and combined dual-resolution (100,50) raw and post-processed wind-speed forecasts with respect to the corresponding pure high-resolution (0,50)
predictions for percentiles 5 (a), 10 (b), 20 (c), 80 (d), 90 (e), and 95 (f) with 95\,\% confidence intervals as functions of the lead time.}
\label{fig:qss_all}
\end{figure}

In Figure \ref{fig:crpss_maes_all}, the CRPSS and MAES  values of raw and locally post-processed wind-speed forecasts with respect to the corresponding pure high-resolution predictions are presented, along with their 95\% bootstrap confidence intervals. In both the CRPSS and MAES panels, it is evident that the pure low-resolution ensemble forecast exhibits far the weakest predictive skill, significantly underperforming the other two ensemble configurations. Similarly, the corresponding local EMOS model performs the worst on average among the post-processed forecasts; however, for longer lead times, the differences between the various EMOS post-processed configurations are not significant. In terms of both investigated scores, there is no lead time where the raw dual-resolution forecast outperforms the pure high-resolution prediction; moreover, it results in significantly negative MAES values for all studied forecast horizons. The only case where the incorporation of low-resolution terms is significantly beneficial is the local EMOS approach; however, even in this case, only up to day 2.

\begin{figure}[th!]
\begin{center}
\epsfig{file=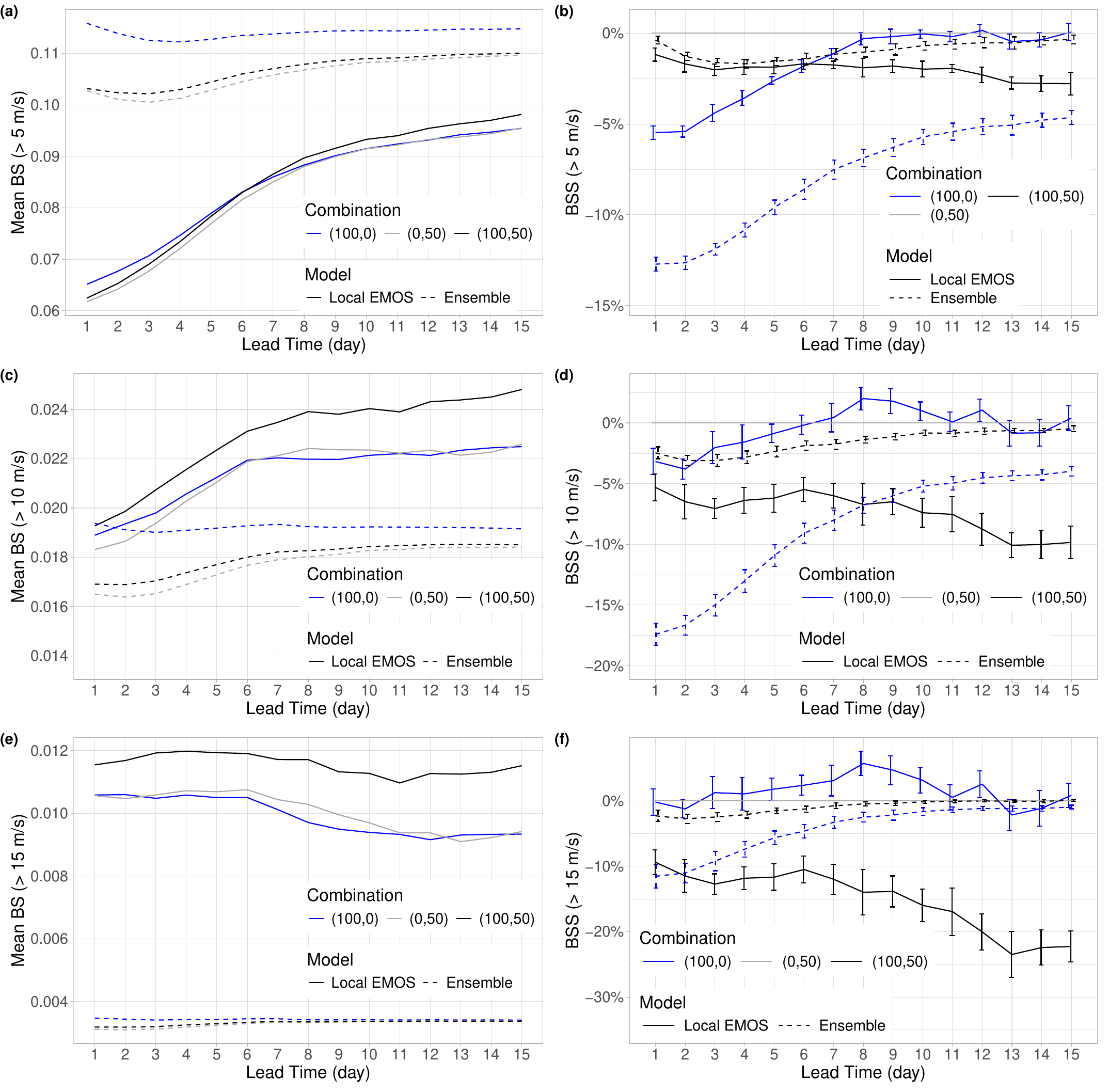, width=\textwidth}
\end{center}
\caption{Mean BS of pure low-resolution (100,0), pure high-resolution (0,50), and combined dual-resolution (100,50) raw and post-processed wind-speed forecasts (a,c,e) and BSS of pure low-resolution (100,0) and combined dual-resolution (100,50) forecasts with respect to the corresponding pure high-resolution (0,50) predictions with 95\,\% confidence intervals (b,d,f) for thresholds 5 m/s (a,b), 10 m/s (c,d), and 15 m/s (e,f) as functions of the lead time.}
\label{fig:bs_bss_all}
\end{figure}

Figure \ref{fig:qss_all} shows the QSS of raw and post-processed wind-speed forecasts from pure low-resolution (100,0) and combined dual-resolution (100,50) ensembles, relative to the corresponding pure high-resolution (0,50) forecasts for selected percentiles, with 95\% confidence intervals as functions of lead time. The pure low-resolution ensemble forecast exhibits the same behaviour as in Figure \ref{fig:crpss_maes_all}; it significantly underperforms its pure high-resolution counterpart for all lead times and all studied quantiles (note that the QSS for the 50th percentile coincides with the MAES). In contrast, the raw dual-resolution ensemble forecast provides a significant advantage over the (0,50) combination for all lead times in most cases. The improvement is highest at the most extreme quantiles and gradually decreases towards the center, completely fading for the 80th percentile (and for the 50th percentile as well, see Figure \ref{fig:crpss_maes_all}) and after day 2 for the 90th percentile. A different situation can be observed in the case of EMOS post-processed predictions; however, the shapes of the corresponding QSS curves are mostly in line with the matching graphs of Figure \ref{fig:crpss_maes_all}. For short lead times, the pure low-resolution EMOS forecast is significantly behind the reference pure high-resolution-based EMOS model but gradually catches up with the increase of the forecast horizon. In general, the deviations are the largest in the central percentiles, where the corresponding QSS fails to be significantly negative only at the longest lead times. Finally, dual-resolution post-processed forecasts result in significantly positive QSS only for the first 2-3 days and only for the larger percentiles.

Figure \ref{fig:bs_bss_all} depicts the mean Brier scores (BS) along with the matching skill scores for the previously investigated raw and post-processed forecasts with respect to the corresponding high-resolution (0,50) predictions. The mean BS values (panels a, c, and e) demonstrate that the advantage of post-processing diminishes progressively with increasing wind-speed thresholds. This indicates that while the truncated normal EMOS model generally performs well, for high wind-speed values, EMOS models utilizing distributions with heavier tails such as log-normal \citep{bl15} or truncated generalized extreme value \citep{bszsz21} can be more suitable. At the lowest threshold of 5 m/s, the superiority of the post-processed models is highly pronounced. For shorter lead times, EMOS models that also build on high-resolution forecasts are superior; however, as the forecast horizon increases, the purely low-resolution forecast begins to catch up and, after day seven, outperforms the (100,50) mixture. According to panels b, d, and f, while for higher wind-speed thresholds, the post-processed pure low-resolution (100,0) forecasts show the highest average advantage, this benefit is only significant for certain forecast lead times (from day 8 to 11). In contrast, the dual-resolution (100,50) forecasts exhibit significantly lower skill across all examined prediction horizons and thresholds, and their disadvantage becomes increasingly pronounced as the threshold increases. A different pattern can be observed for the raw forecasts: here, the dual-resolution (100,50) configurations exhibit a performance more comparable to that of the high-resolution (0,50) forecasts. However, neither configuration shows a significant advantage over pure high-resolution forecasts at any forecast lead time or threshold.

\begin{figure}[t]
\begin{center}
\epsfig{file=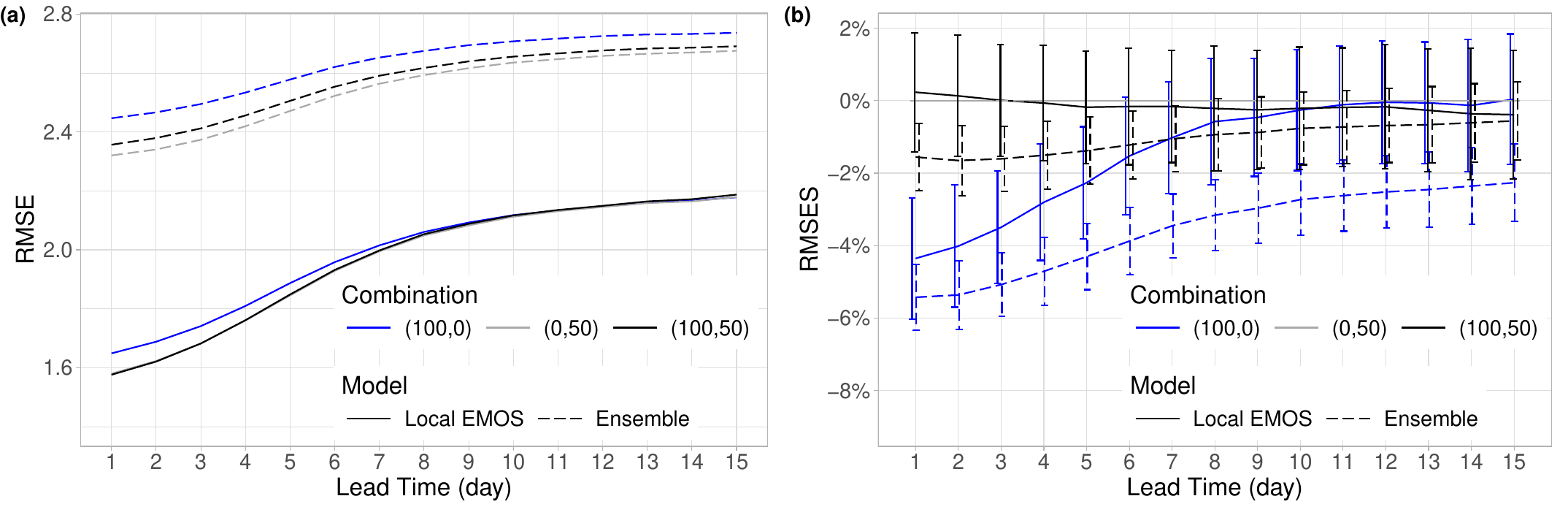, width=\textwidth}
\end{center}
\caption{RMSE of the mean of pure low-resolution (100,0), pure high-resolution (0,50), and combined dual-resolution (100,50) raw and post-processed wind-speed forecasts (a) and RMSES with respect to the corresponding pure high-resolution predictions with 95\,\% confidence intervals (b) as functions of the lead time.}
\label{fig:rmse_rmses_all}
\end{figure}

Finally, the RMSE values of Figure \ref{fig:rmse_rmses_all}a again confirm the general advantage of statistical post-processing. Here, all EMOS models clearly perform better than the raw forecasts, especially those based on dual-resolution and pure high-resolution ensembles. However, as the forecast lead time increases, their performance becomes similar. Furthermore, according to the skill scores of Figure \ref{fig:rmse_rmses_all}b, none of the forecasts does significantly better than the corresponding pure high-resolution reference.

\subsection{Forecast mixtures}
\label{subs4.2}

In the following, our goal is to investigate how the inclusion of high-resolution ($\text{T}_{\text{CO}}1279$) ensemble members affects the predictive performance of the forecasts when gradually added to a base of 50 low-resolution ($\text{T}_{\text{CO}}319$) predictions. We separately assess the forecast skill of the raw ensemble (Section \ref{subs4.2.1}) and the locally calibrated EMOS models (Section \ref{subs4.2.2}); the latter choice of the training data selection method is based on the findings of Section \ref{subs4.1}. Specifically, we assess the predictive performance of the raw and post-processed ECMWF wind-speed forecasts by gradually adding 1, 2, 4, 8, 16, and 32 high-resolution  ensemble members to the original set of 50 low-resolution forecasts. 

\subsubsection{Ensemble predictions}
\label{subs4.2.1}

\begin{figure}[t]
\begin{center}
\epsfig{file=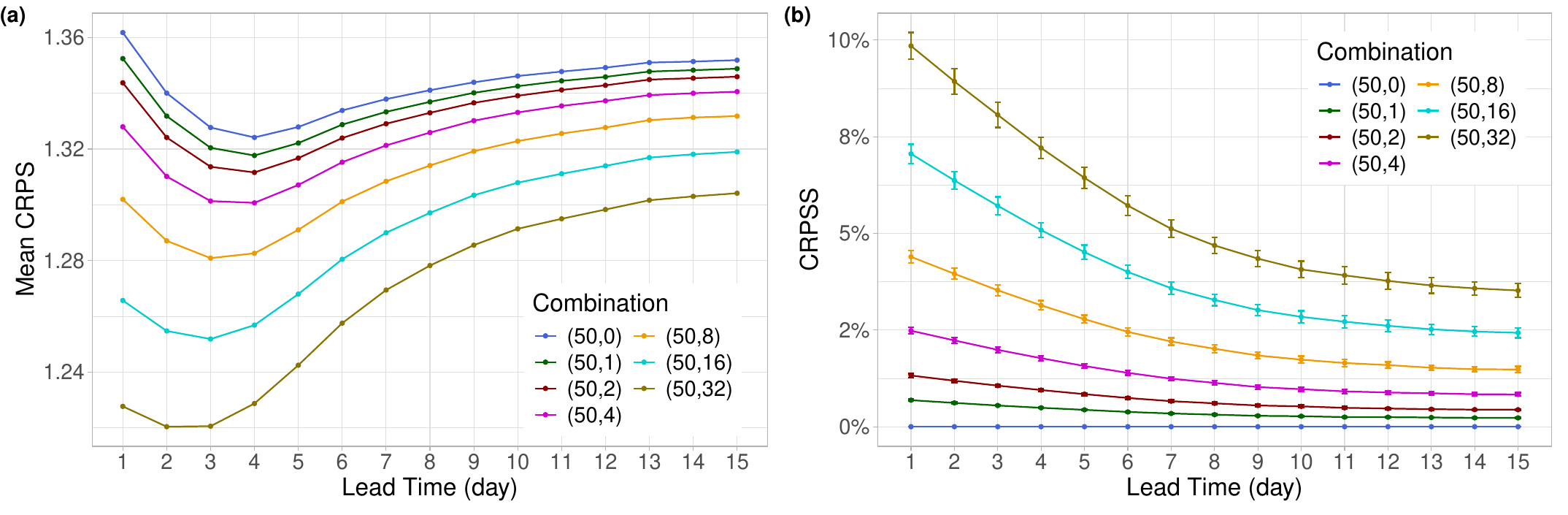, width=\textwidth}
\end{center}
\caption{Mean CRPS of various combinations of low- and high-resolution raw wind-speed ensemble forecasts (a) and CRPSS of mixtures containing high-resolution members with respect to the pure low-resolution (50,0) prediction with 95\,\% confidence intervals (b) as functions of the lead time.}
\label{fig:crps_crpss_ens}
\end{figure}

Figure \ref{fig:crps_crpss_ens} illustrates the predictive accuracy of these combined ensembles. Panel (a) presents the mean CRPS across lead times, while panel (b) shows the CRPSS relative to the baseline (50,0) low-resolution ensemble. The inclusion of high-resolution members results in consistent improvements in forecast skill, particularly at short to medium lead times. The greatest improvement—nearly 10\% CRPSS -- is achieved by the (50,32) configuration at day 1, followed by (50,16) with gains of up to 7\%, and (50,8) with approximately 5\%. Even configurations with only one or two additional high-resolution members, such as (50,1) or (50,2), yield slight improvements. While the (50,32) setup provides the most substantial overall benefit relative to the reference, the magnitude of this advantage decreases with increasing forecast lead time across all model configurations; nevertheless, the improvement remains statistically significant for all lead times and ensemble configurations, as confirmed by 95\% bootstrap confidence intervals. 

\begin{figure}[t!]
\begin{center}
\epsfig{file=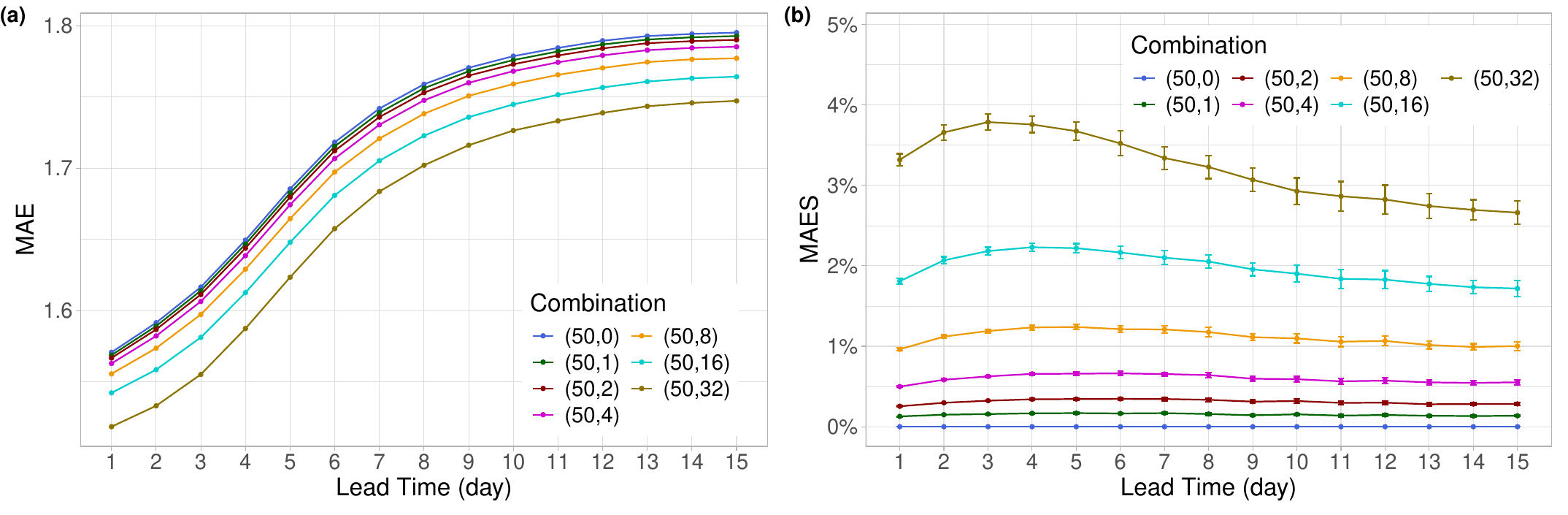, width=\textwidth}
\end{center}
\caption{MAE of the medians of various combinations of low- and high-resolution raw wind-speed ensemble forecasts (a) and MAES of mixtures containing high-resolution members with respect to the pure low-resolution (50,0) prediction with 95\,\% confidence intervals (b) as functions of the lead time.}
\label{fig:mae_maes_ens}
\end{figure}

In terms of model ranking, a similar pattern can be observed in Figure \ref{fig:mae_maes_ens} when analyzing the median of the forecasts: as more members are added to the ensemble consisting solely of low-resolution forecasts, the skill consistently improves. Moreover, it appears that for all model configurations, the positive impact of high-resolution members is most pronounced at days 3 and 4, and this advantage becomes increasingly evident as more high-resolution members are included. However, unlike with the CRPS, increasing lead time does not necessarily correspond to a decreasing advantage relative to the reference forecast here. This highlights that the benefit of adding high-resolution ensemble members is particularly clear in the median forecasts, further underscoring their value in enhancing predictive accuracy, especially at these intermediate lead times. 

\begin{figure}[t!]
\begin{center}
\epsfig{file=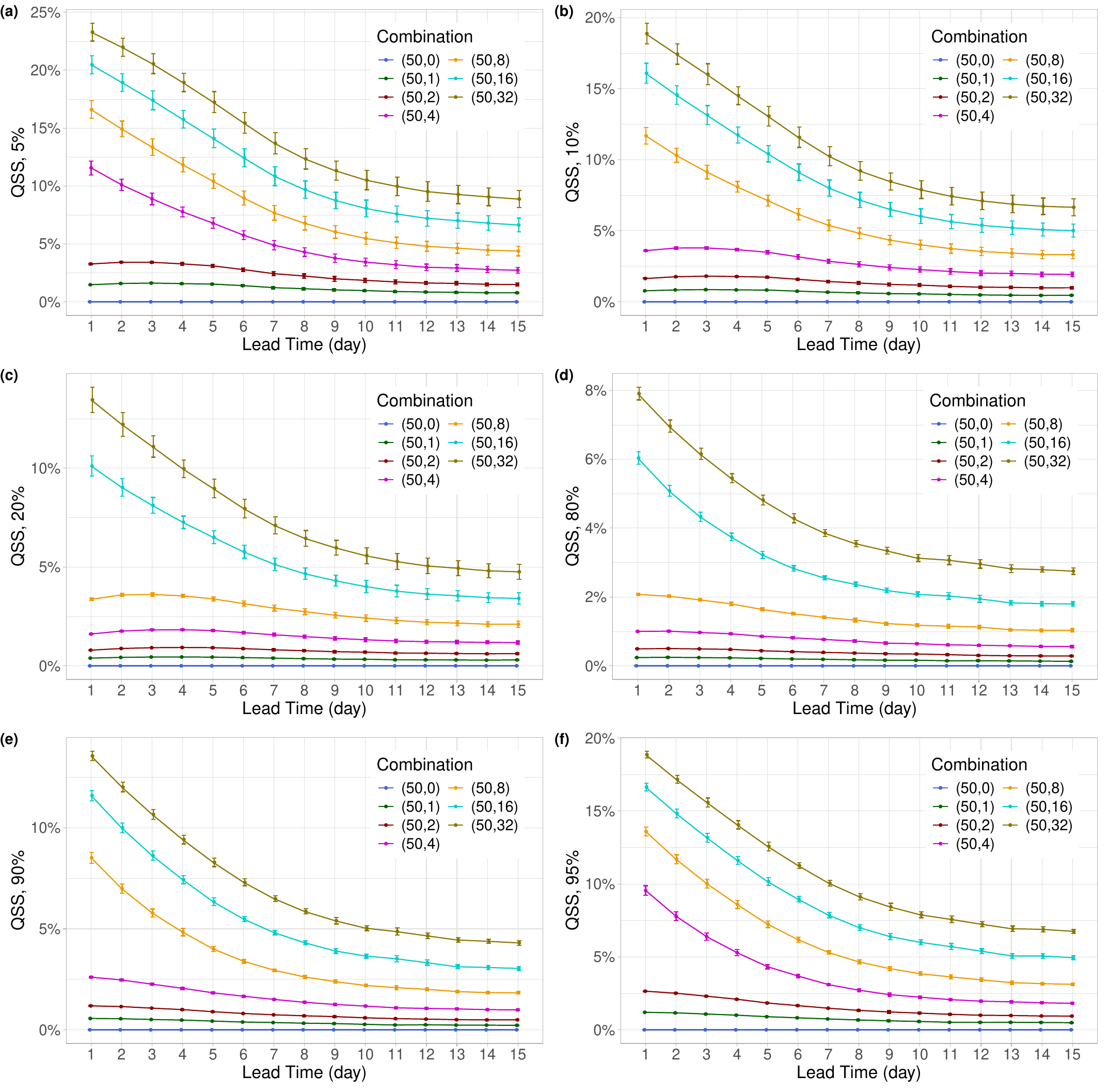, width=\textwidth}
\end{center}
\caption{QSS with respect to the pure low-resolution (50,0) ensemble prediction of wind-speed for percentiles 5 (a), 10 (b), 20 (c), 80 (d), 90 (e), and 95 (f) of mixtures containing high-resolution members with 95\,\% confidence intervals as functions of the lead time.}
\label{fig:qss_ens}
\end{figure}

\begin{figure}[h!]
\begin{center}
\epsfig{file=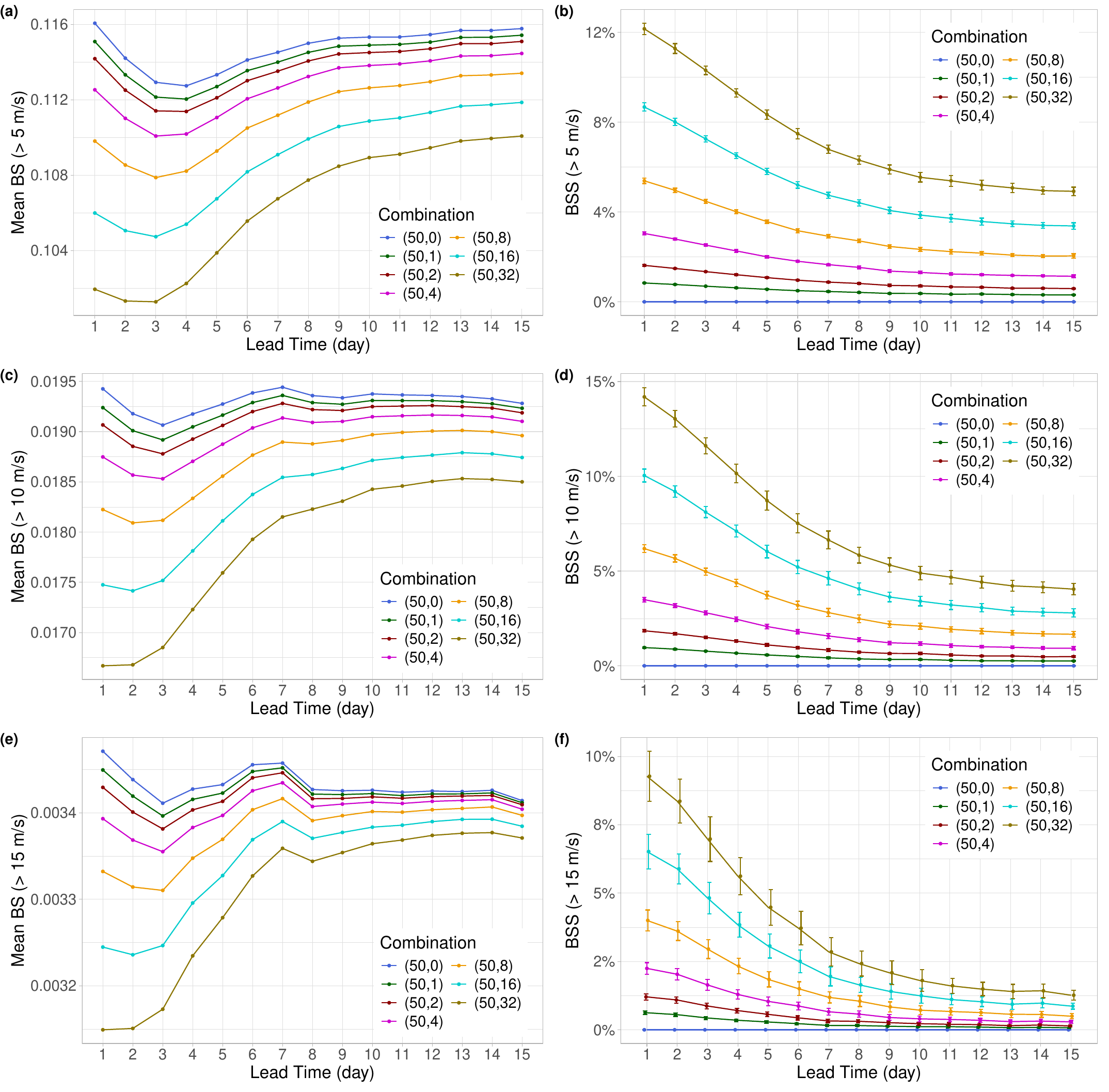, width=\textwidth}
\end{center}
\caption{Mean BS of various combinations of raw low- and high-resolution wind-speed forecasts (a,c,e) and BSS of mixtures containing high-resolution members with respect to the pure low-resolution (50,0) prediction with 95\,\% confidence intervals (b,d,f) for thresholds 5 m/s (a,b), 10 m/s (c,d), and 15 m/s (e,f) as functions of the lead time.}
\label{fig:bs_bss_ens}
\end{figure}

Figure \ref{fig:qss_ens} shows various quantile skill scores for all previously considered ensemble configurations containing high-resolution members with respect to the pure low-resolution forecast. The overall ranking of the models remains consistent with that observed in Figures \ref{fig:crps_crpss_ens} and \ref{fig:mae_maes_ens}. However, the magnitude of the improvement varies across different percentiles. For the more extreme percentiles (5\% and 95\%), forecasts using only four high-resolution members show a notably greater advantage compared to other percentiles, closely followed by those incorporating 8, 16, and 32 high-resolution members. More generally, we observe that symmetrically positioned percentiles (such as 20\% – 80\% and 10\% – 90\%) exhibit similar trends in model performance as more high-resolution members are added, though the magnitude of the improvements differs. For instance, at the 20\% and 80\% percentiles, the largest gains are achieved by configurations including 16 or 32 high-resolution members, whereas at the 10\% and 90\% percentiles, the model using 8 high-resolution members also ranks among the top performers. 

Figure \ref{fig:bs_bss_ens} reporting the Brier scores for the events of wind-speed exceeding thresholds 5, 10, and 15 m/s for different raw forecast mixtures does not change the general picture. The overall trend is similar across all three thresholds: the more high-resolution members are added to the 50 low-resolution members, the better the predictive performance becomes. However, for all thresholds, these improvements become less pronounced as the forecast lead time increases. Based on panels b, d, and f of Figure \ref{fig:bs_bss_ens}, significant differences in forecast performance can be observed between the ensemble configurations; however, for higher wind-speed thresholds, the confidence intervals become increasingly wide, indicating greater uncertainty.

\begin{figure}[t]
\begin{center}
\epsfig{file=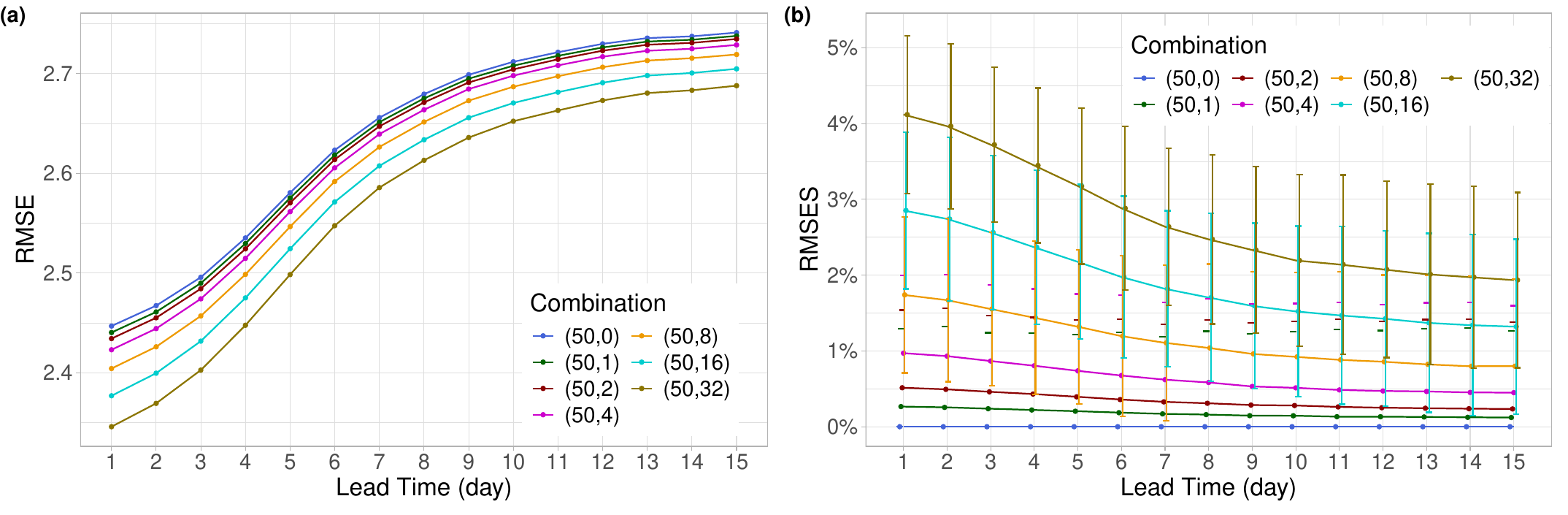, width=\textwidth}
\end{center}
\caption{RMSE of the means of various combinations of low- and high-resolution raw wind-speed ensemble forecasts (a) and RMSES of mixtures containing high-resolution members with respect to the pure low-resolution (50,0) prediction with 95\,\% confidence intervals (b) as functions of the lead time.}
\label{fig:rmse_rmses_ens}
\end{figure}

\begin{figure}[t]
\begin{center}
\epsfig{file=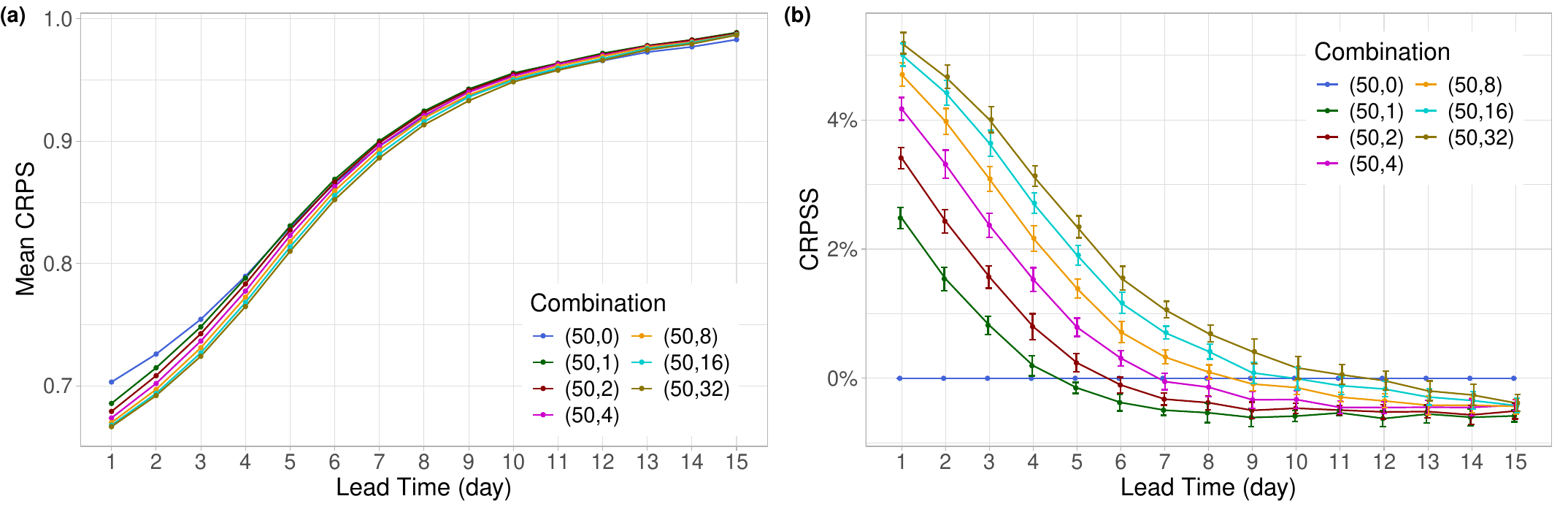, width=\textwidth}
\end{center}
\caption{Mean CRPS of various combinations of locally post-processed low- and high-resolution wind-speed forecasts (a) and CRPSS of mixtures containing high-resolution members with respect to the pure low-resolution (50,0) prediction with 95\,\% confidence intervals (b) as functions of the lead time.}
\label{fig:crps_crpss_loc}
\end{figure}

\begin{figure}[h!]
\begin{center}
\epsfig{file=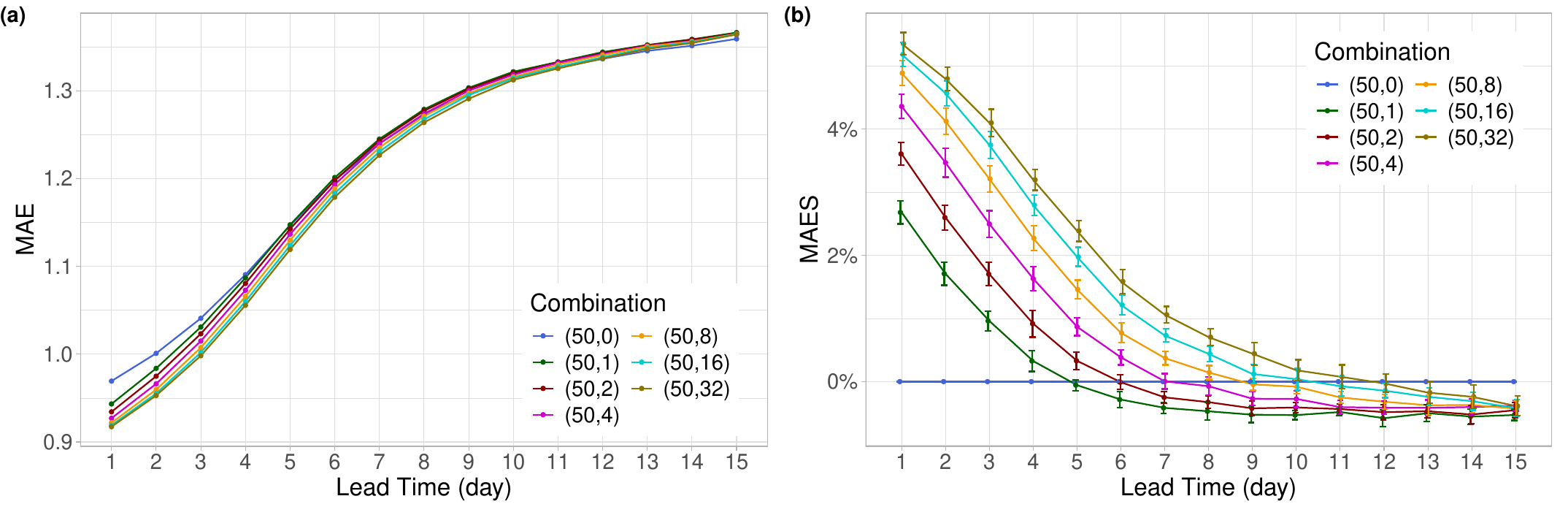, width=\textwidth}
\end{center}
\caption{MAE of the medians of various combinations of locally post-processed low- and high-resolution wind-speed forecasts (a) and MAES of mixtures containing high-resolution members with respect to the pure low-resolution (50,0) prediction with 95\,\% confidence intervals (b) as functions of the lead time.}
\label{fig:mae_maes_loc}
\end{figure}

Finally, Figure \ref{fig:rmse_rmses_ens} presents the mean RMSE and the corresponding skill score values. While the overall model ranking is similar to that shown in Figure \ref{fig:mae_maes_ens}, the confidence intervals suggest that including more high-resolution forecasts does not necessarily result in statistically significant improvements compared to configurations with fewer high-resolution members. Furthermore, the benefit of adding high-resolution members again diminishes with increasing forecast lead time, a pattern that is most pronounced for forecasts using a larger number of high-resolution members.

\subsubsection{Post-processed forecasts}
\label{subs4.2.2}
In the following, we evaluate the predictive performance of the forecast mixtures introduced in the previous section, post-processed using the local EMOS method, as this model performed best for the operational forecasts evaluated in Section \ref{subs4.1}.

Figure \ref{fig:crps_crpss_loc}a displays the mean CRPS values for the different post-processed forecast combinations. Similar to Figure \ref{fig:crps_crpss_ens}a, the more high-resolution members are included in the post-processing, the better the predictive performance. Furthermore, differences between the models gradually diminish as the forecast lead time increases, indicating that the benefit of including further high-resolution forecasts becomes negligible at longer lead times. One should also note that post-processing substantially reduces the deviations of the various combinations, which is completely in line with the findings of \citet{blszb19} (see also Figure \ref{fig:crps_crpss_all}a). Finally, the skill scores of Figure \ref{fig:crps_crpss_loc}b reveal that all models utilizing high-resolution predictions significantly outperform the reference forecast based solely on low-resolution members up to day 4. Beyond this point, the inclusion of additional high-resolution members helps maintain this performance advantage over longer lead times, with the (50,32) combination showing the greatest overall benefit. These conclusions are further supported by Figure \ref{fig:mae_maes_loc}, which presents the MAE values and corresponding skill scores for the same post-processed forecast combinations.

\begin{figure}[h!]
\begin{center}
\epsfig{file=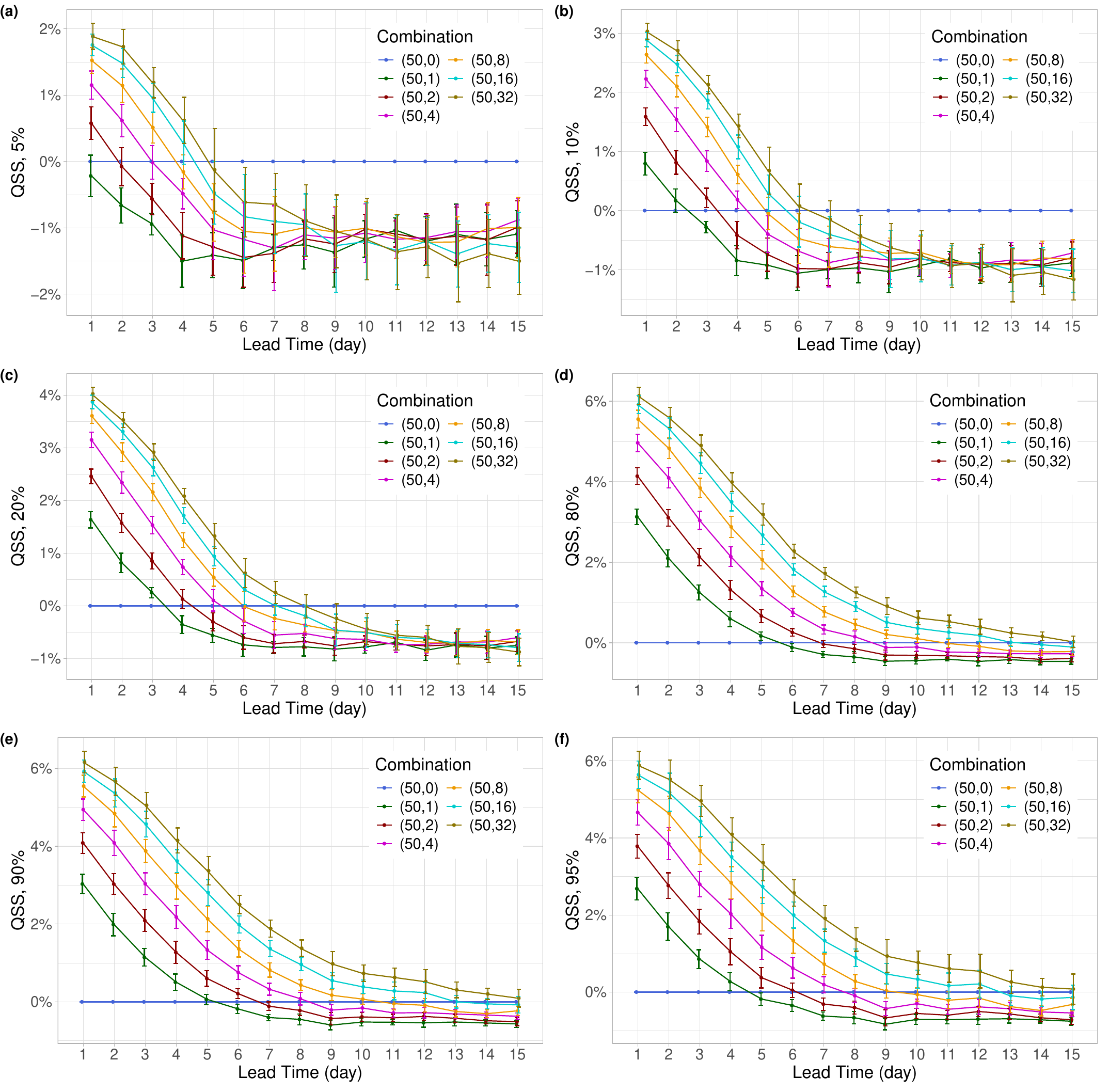, width=\textwidth}
\end{center}
\caption{QSS with respect to the locally post-processed pure low-resolution (50,0) prediction of wind-speed for percentiles 5 (a), 10 (b), 20 (c), 80 (d), 90 (e), and 95 (f) of post-processed mixtures containing high-resolution members with 95\,\% confidence intervals as functions of the lead time.}
\label{fig:qss_loc}
\end{figure}

We also examined the performance of the post-processed models based on the QSS, as shown in Figure \ref{fig:qss_loc}. The overall behaviour of the models is consistent with previous findings: the inclusion of more high-resolution members in the post-processing generally leads to greater improvements over the pure low-resolution reference forecasts. However, the magnitude of this improvement varies across different percentiles. For lower percentiles, the benefit of including high-resolution members is smaller. In an extreme case -- at the 5th percentile -- the (50,1) combination even performs significantly worse than the pure low-resolution reference. In contrast, the performance advantage of all models increases consistently with higher percentiles.

\begin{figure}[h!]
\begin{center}
\epsfig{file=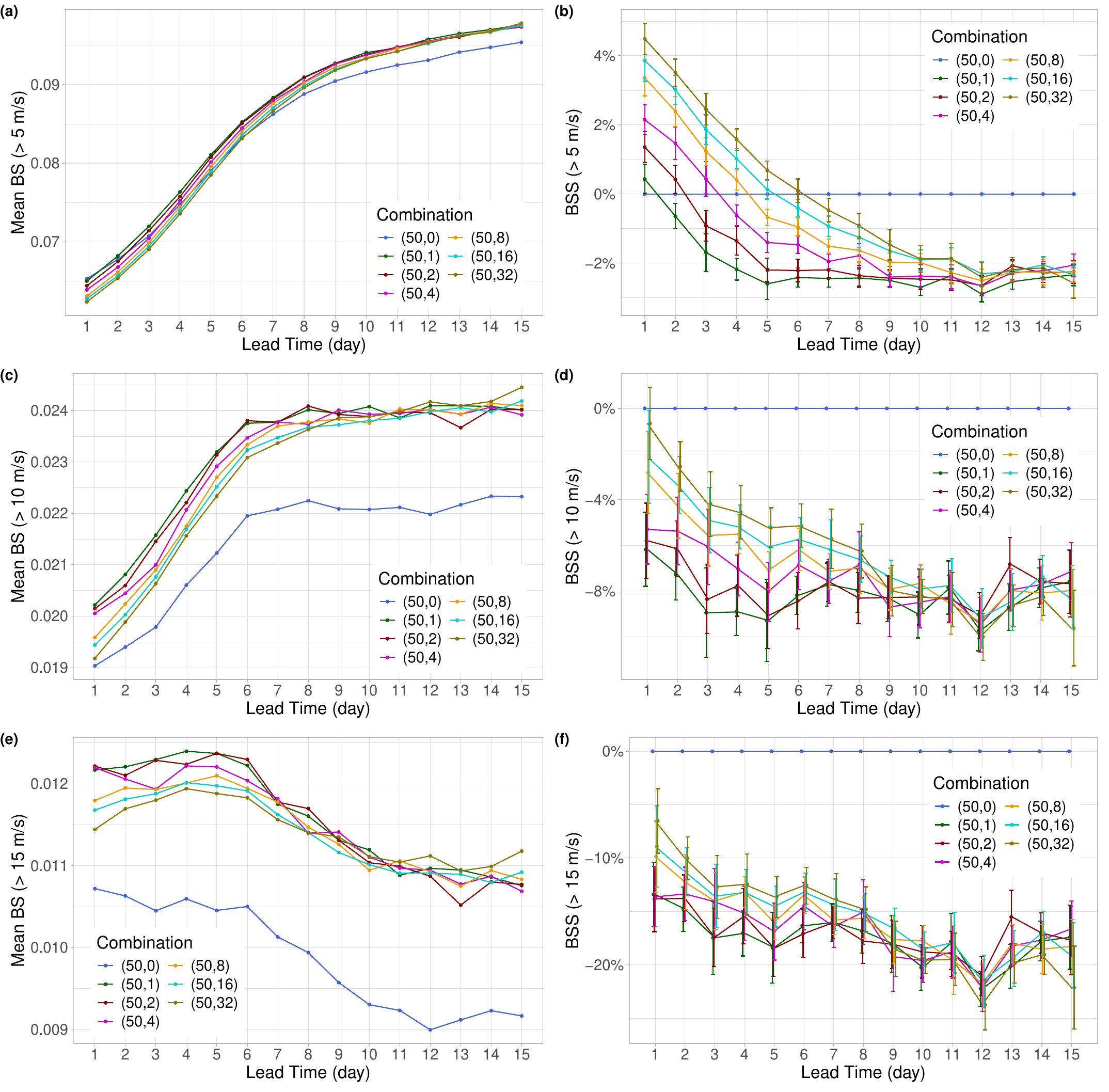, width=\textwidth}
\end{center}
\caption{Mean BS of various combinations of locally post-processed low- and high-resolution wind-speed forecasts (a,c,e) and BSS of mixtures containing high-resolution members with respect to the pure low-resolution (50,0) prediction with 95\,\% confidence intervals (b,d,f) for thresholds 5 m/s (a,b), 10 m/s (c,d), and 15 m/s (e,f) as functions of the lead time.}
\label{fig:bs_bss_loc}
\end{figure}

Figure \ref{fig:bs_bss_loc} shows the mean Brier scores corresponding to the locally post-processed forecast combinations for the low (5 m/s), moderate (10 m/s), and high (15 m/s) wind-speed thresholds, and the matching skill scores with respect to the EMOS model based on the pure low-resolution (50,0) prediction. For low wind-speeds, the patterns are similar to those observed in the previous CRPS and MAE figures. In general, including more ensemble members leads to better model performance, though, as before, this advantage holds only up to a certain forecast horizon. At low wind-speeds, the benefit of the pure low-resolution forecast is now even more pronounced (Figure \ref{fig:bs_bss_loc}a), and according to Figure \ref{fig:bs_bss_loc}b, the advantage of models incorporating high-resolution members remains significant only up to day 5 at best. For the moderate and high wind-speed thresholds, the picture changes more drastically: none of the post-processed models show a statistically significant improvement over the pure low-resolution (50,0) reference.

\begin{figure}[t]
\begin{center}
\epsfig{file=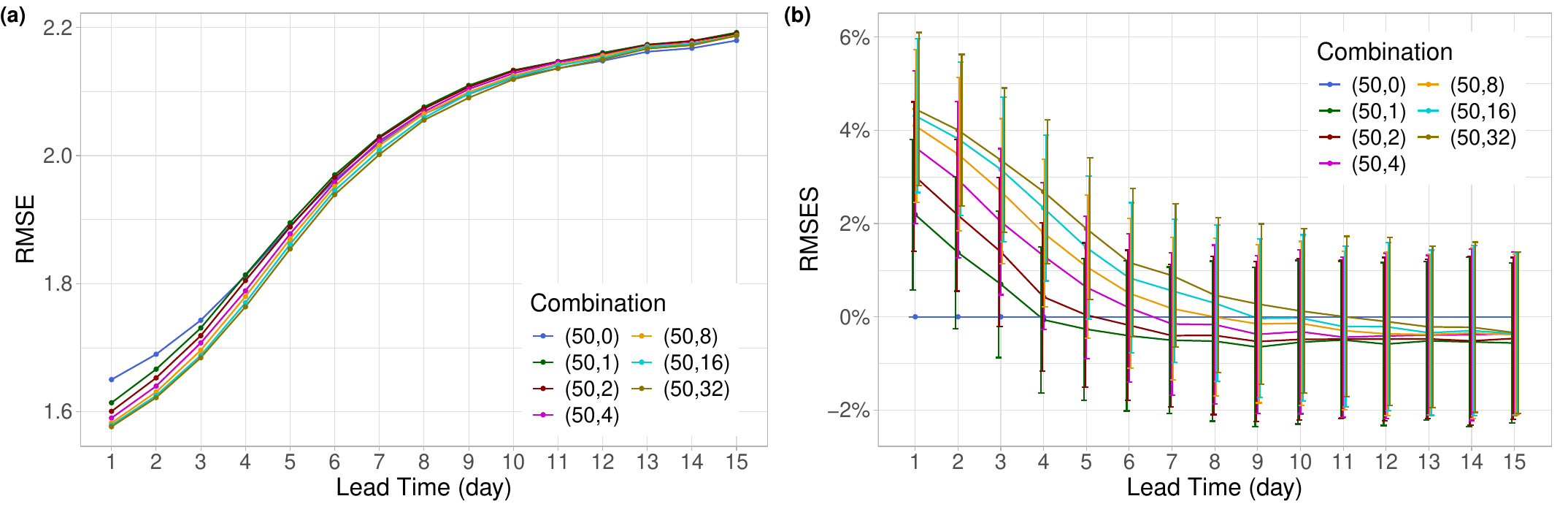, width=\textwidth}
\end{center}
\caption{RMSE of the means of various combinations of locally post-processed low- and high-resolution wind-speed forecasts (a) and RMSES of mixtures containing high-resolution members with respect to the pure low-resolution (50,0) prediction with 95\,\% confidence intervals (b) as functions of the lead time.}
\label{fig:rmse_rmses_loc}
\end{figure}

Finally, Figure \ref{fig:rmse_rmses_loc}, which presents the RMSE and related skill scores, further confirms the above findings. However, the overlapping 95\% confidence intervals among the models suggest greater uncertainty regarding the significance of the performance differences with respect to the pure low-resolution (50,0) prediction.

\section{Conclusions}
\label{sec5}

We investigate the forecast skill of raw and post-processed 50-member medium-range and 100-member extended-range forecasts of 10-m wind-speed up to day 15 at $\text{T}_{\text{CO}}1279$ (high) and $\text{T}_{\text{CO}}319$ (low) resolutions, respectively, and their various combinations. This study demonstrates that among the raw ECMWF ensemble forecasts, the low-resolution (100,0) configuration consistently shows the weakest predictive performance for all investigated verification measures, while the high-resolution (0,50) and dual-resolution (100,50) setups perform similarly. Nevertheless, in most cases, the former maintains a slight but significant edge, especially for short lead times; the dual-resolution forecast is significantly superior to the single-resolution predictions only in terms of the mean quantile scores for the low and very high percentiles.

Post-processing -- particularly with local and semi-local EMOS models -- offers substantial improvements in forecast skill in terms of all studied evaluation metrics, but the Brier scores for medium and high thresholds. The local method demonstrates the best overall performance, especially during the early forecast days. Moreover, when beneficial, statistical calibration considerably reduces the differences between the various forecast setups. In contrast to the raw ensemble predictions, the pure low-resolution EMOS forecast starts off lagging significantly behind the high-resolution EMOS benchmark but improves gradually over time, slowly closing the performance gap. Furthermore, for very short lead times, the use of EMOS post-processed dual-resolution forecasts seems to be advantageous; it is significantly superior to the single-resolution setups in terms of the mean CRPS and mean QS for percentiles not less than the median.

The above findings suggest that spatial resolution is superior to the ensemble size; augmenting a sufficiently large ensemble of high-resolution forecasts with low-resolution predictions does not necessarily result in a gain in forecast skill. However, our study also highlights the clear benefit of the other direction, namely, incorporating high-resolution members into low-resolution ensemble forecasts. 
Again, based on multiple verification scores, the addition of high-resolution members consistently enhances the predictive performance of the raw ECMWF wind-speed forecasts for all studied forecast horizons. The most significant gains are observed in configurations with the highest number of high-resolution members, though even minimal inclusion yields slight improvements. After post-processing, the advantage of dual-resolution modelling depends on the number of incorporated high-resolution predictions. In general, the more high-resolution members are involved, the longer the lead time until the superiority of the dual-resolution combination over the EMOS model relying merely on the low-resolution ensemble is significant. 

The results of the current study suggest several avenues for further research. For post-processing, we utilized a very simple univariate approach incorporating neither spatial dependencies between the SYNOP stations nor temporal dependencies between the different forecast horizons. Thus, on the one hand, one can investigate the performance of more sophisticated state-of-the-art post-processing methods such as the parametric machine-learning-based DRN \citep{rl18} or versions of the non-parametric quantile regression \citep[see e.g.][]{brem20,songetal24} in the dual-resolution setup. On the other hand, the study of multivariate post-processing approaches -- classical two-step methods such as the ensemble copula coupling \citep{sch13} or the Schaake shuffle \citep{clarketal04} or data-driven techniques such as generative adversarial networks \citep{dh21} or scoring-rule-based generative models \citep{cjsl24} -- can also lead us to a better understanding of the effects of mixing high- and low-resolution ensemble members. Finally, the increasing popularity of data-driven weather forecasts and the launch of the ensemble version of the ECMWF's Artificial Intelligence Forecasting System \citep[AIFS-CRPS;][]{lang2024aifs-crps} currently issued at a 28 km grid resolution, naturally induces the question of whether mixing IFS and AIFS-CRPS ensemble predictions results in improved predictive performance.

\bigskip
\noindent
{\bf Acknowledgments.} \ This work was supported by the EK{\"O}P-24-3-II University Research Scholarship Program of the Ministry for Culture and Innovation, funded by the National Research, Development and Innovation Fund.  The authors also gratefully acknowledge the support of the  National Research, Development, and Innovation Office under Grant No. K142849. Finally, they are indebted to Martin Leutbecher for his suggestions and for providing the ECMWF dual-resolution wind-speed data.

\bibliographystyle{apalike}
\bibliography{references}

\begin{thebibliography}{}

\bibitem[Baran and Lakatos, 2024]{bl24}
Baran, S. and Lakatos, M. (2024).
\newblock Clustering-based spatial interpolation of parametric postprocessing
  models.
\newblock {\em Weather Forecast.}, \textbf{39}(11):1591--1604.

\bibitem[Baran and Lerch, 2015]{bl15}
Baran, S. and Lerch, S. (2015).
\newblock Log-normal distribution based emos models for probabilistic wind
  speed forecasting.
\newblock {\em Q. J. R. Meteorol. Soc.}, \textbf{141}(691):2289--2299.

\bibitem[Baran et~al., 2019]{blszb19}
Baran, S., Leutbecher, M., Szab{\'o}, M., and Ben~Bouall{\`e}gue, Z. (2019).
\newblock Statistical post-processing of dual-resolution ensemble forecasts.
\newblock {\em Q. J. R. Meteorol. Soc.}, \textbf{145}(721):1705--1720.

\bibitem[Baran et~al., 2021]{bszsz21}
Baran, S., Szokol, P., and Szab{\'o}, M. (2021).
\newblock Truncated generalized extreme value distribution-based ensemble model
  output statistics model for calibration of wind speed ensemble forecasts.
\newblock {\em Environmetrics}, \textbf{32}(6):paper e2678.

\bibitem[{Ben Bouall{\`e}gue} et~al., 2020]{bbhwhr20}
{Ben Bouall{\`e}gue}, Z., Haiden, T., Weber, N.~J., Hamill, T.~M., and
  Richardson, D.~S. (2020).
\newblock Accounting for representativeness in the verification of ensemble
  precipitation forecasts.
\newblock {\em Mon. Weather Rev.}, \textbf{148}(5):2049--2062.

\bibitem[Bentzien and Friederichs, 2014]{bf14}
Bentzien, S. and Friederichs, P. (2014).
\newblock Decomposition and graphical portrayal of the quantile score.
\newblock {\em Q. J. R. Meteorol. Soc.}, \textbf{140}(683):1924--1934.

\bibitem[Bremnes, 2020]{brem20}
Bremnes, J.~B. (2020).
\newblock Ensemble postprocessing using quantile function regression based on
  neural networks and {Bernstein} polynomials.
\newblock {\em Mon. Wea. Rev.}, \textbf{148}(1):403--414.

\bibitem[Buizza, 2018]{buizza18}
Buizza, R. (2018).
\newblock Ensemble forecasting and the need for calibration.
\newblock In Vannitsem, S., Wilks, D.~S., and Messner, J.~W., editors, {\em
  {Statistical Postprocessing of Ensemble Forecasts}}, pages 15--48. Elsevier.

\bibitem[Chen et~al., 2024]{cjsl24}
Chen, J., Janke, T., Steinke, F., and Lerch, S. (2024).
\newblock Generative machine learning methods for multivariate ensemble
  postprocessing.
\newblock {\em Ann. Appl. Stat.}, \textbf{18}(1):159--189.

\bibitem[Clark et~al., 2004]{clarketal04}
Clark, M., Gangopadhyay, S., Hay, L., Rajagopalan, B., and Wilby, R. (2004).
\newblock The schaake shuffle: A method for reconstructing space–time
  variability in forecasted precipitation and temperature fields.
\newblock {\em J. Hydrometeorol.}, \textbf{5}(1):243--262.

\bibitem[Dai and Hemri, 2021]{dh21}
Dai, Y. and Hemri, S. (2021).
\newblock Spatially coherent postprocessing of cloud cover ensemble forecasts.
\newblock {\em Mon. Weather Rev.}, \textbf{149}(12):3923--3937.

\bibitem[ECMWF, 2024]{ecmwf24}
ECMWF (2024).
\newblock {\em IFS Documentation CY49R1 -- Part V: Ensemble Prediction System}.
\newblock ECMWF, Reading.

\bibitem[Gasc\'on et~al., 2019]{glhrblp19}
Gasc\'on, E., Lavers, D., Hamill, T.~M., Richardson, D.~S., {Ben Bouall\`egue},
  Z., Leutbecher, M., and Pappenberger, F. (2019).
\newblock Statistical postprocessing of dual-resolution ensemble precipitation
  forecasts across {Europe}.
\newblock {\em Q. J. R. Meteorol. Soc.}, \textbf{145}(724):3218--3235.

\bibitem[Gneiting, 2011]{gneiting11}
Gneiting, T. (2011).
\newblock Making and evaluating point forecasts.
\newblock {\em J. Amer. Statist. Assoc.}, \textbf{106}(494):746--762.

\bibitem[Gneiting and Raftery, 2007]{gr07}
Gneiting, T. and Raftery, A.~E. (2007).
\newblock Strictly proper scoring rules, prediction, and estimation.
\newblock {\em J. Am. Stat. Assoc.}, \textbf{102}(477):359--378.

\bibitem[Gneiting et~al., 2005]{grgw}
Gneiting, T., Raftery, A.~E., Westveld, A.~H., and Goldman, T. (2005).
\newblock {Calibrated probabilistic forecasting using ensemble model output
  statistics and minimum {CRPS} estimation}.
\newblock {\em Mon. Weather Rev.}, {\textbf{133}}(5):1098--1118.

\bibitem[Hemri et~al., 2014]{hemri14}
Hemri, S., Scheuerer, M., Pappenberger, F., Bogner, K., and Haiden, T. (2014).
\newblock Trends in the predictive performance of raw ensemble weather
  forecasts.
\newblock {\em Geophys. Res. Lett.}, \textbf{41}(24):9197--9205.

\bibitem[Jordan et~al., 2019]{jkl19}
Jordan, A., Krüger, F., and Lerch, S. (2019).
\newblock Evaluating probabilistic forecasts with {scoringRules}.
\newblock {\em J. Stat. Softw.}, \textbf{90}(12):1--37.

\bibitem[Lang et~al., 2024]{lang2024aifs-crps}
Lang, S., Alexe, M., Clare, M. C.~A., Roberts, C., Adewoyin, R.,
  Bouall{\`e}gue, Z.~B., Chantry, M., Dramsch, J., Dueben, P.~D., Hahner, S.,
  Maciel, P., Prieto-Nemesio, A., O'Brien, C., Pinault, F., Polster, J.,
  Raoult, B., Tietsche, S., and Leutbecher, M. (2024).
\newblock {AIFS-CRPS}: Ensemble forecasting using a model trained with a loss
  function based on the continuous ranked probability score.
\newblock arXiv:2412.15832. \url{https://doi.org/10.48550/arXiv.2412.15832}.

\bibitem[Lerch and Baran, 2017]{lb17}
Lerch, S. and Baran, S. (2017).
\newblock Similarity-based semilocal estimation of post-processing models.
\newblock {\em J. R. Stat. Soc.}, {\bf 66}(1):29--51.

\bibitem[Leutbecher and Ben~Bouall{\`e}gue, 2020]{lb20}
Leutbecher, M. and Ben~Bouall{\`e}gue, Z. (2020).
\newblock On the probabilistic skill of dual-resolution ensemble forecasts.
\newblock {\em Q. J. R. Meteorol. Soc.}, \textbf{146}(727):707--723.

\bibitem[Murphy, 1973]{murphy73}
Murphy, A.~H. (1973).
\newblock Hedging and skill scores for probability forecasts.
\newblock {\em J. Appl. Meteorol.}, \textbf{12}(1):215--223.

\bibitem[Politis and Romano, 1994]{pr94}
Politis, D.~N. and Romano, J.~P. (1994).
\newblock The stationary bootstrap.
\newblock {\em J. Am. Stat. Assoc.}, \textbf{89}(428):1303--1313.

\bibitem[Rasp and Lerch, 2018]{rl18}
Rasp, S. and Lerch, S. (2018).
\newblock Neural networks for postprocessing ensemble weather forecasts.
\newblock {\em Mon. Weather Rev.}, \textbf{146}(11):3885--3900.

\bibitem[Schefzik et~al., 2013]{sch13}
Schefzik, R., Thorarinsdottir, T.~L., and Gneiting, T. (2013).
\newblock Uncertainty quantification in complex simulation models using
  ensemble copula coupling.
\newblock {\em Statist.\ Sci.}, \textbf{28}(4):616--640.

\bibitem[Song et~al., 2024]{songetal24}
Song, M., Yang, D., Lerch, S., Xia, X., Yagli, G.~M., Bright, J.~M., Shen, Y.,
  Liu, B., and Liu, Xingli~Mayer, M.~J. (2024).
\newblock Non-crossing quantile regression neural network as a calibration tool
  for ensemble weather forecasts.
\newblock {\em Adv. Atmos. Sci.}, \textbf{41}(7):1417--1437.

\bibitem[Szab{\'o} et~al., 2023]{szgb23}
Szab{\'o}, M., Gasc{\'o}n, E., and Baran, S. (2023).
\newblock Parametric postprocessing of dual-resolution precipitation forecasts.
\newblock {\em Weather Forecast.}, \textbf{38}(8):1313--1322.

\bibitem[Taillardat, 2021]{tail21}
Taillardat, M. (2021).
\newblock Skewed and mixture of {Gaussian} distributions for ensemble
  postprocessing.
\newblock {\em Atmosphere}, \textbf{12}(8):paper 966.

\bibitem[Thorarinsdottir and Gneiting, 2010]{tg10}
Thorarinsdottir, T.~L. and Gneiting, T. (2010).
\newblock Probabilistic forecasts of wind speed: Ensemble model output
  statistics by using heteroscedastic censored regression.
\newblock {\em J. Roy. Stat. Soc.}, \textbf{173A}(2):371--388.

\bibitem[Vannitsem et~al., 2021]{vbd21}
Vannitsem, S., Bremnes, J.~B., Demaeyer, J., Evans, G.~R., Flowerdew, J.,
  Hemri, S., Lerch, S., Roberts, N., Theis, S., Atencia, A., {Ben Boual\`egue},
  Z., Bhend, J., Dabernig, M., De~Cruz, L., Hieta, L., Mestre, O., Moret, L.,
  Odak~Plenkovi\v{c}, I., Schmeits, M., Taillardat, M., Van~den Bergh, J.,
  Van~Schaeybroeck, B., Whan, K., and Ylhaisi, J. (2021).
\newblock Statistical postprocessing for weather forecasts -- review,
  challenges and avenues in a big data world.
\newblock {\em Bull. Amer. Meteorol. Soc.}, \textbf{102}(3):E681--E699.

\bibitem[Veldkamp et~al., 2021]{vwds21}
Veldkamp, S., Whan, K., Dirksen, S., and Schmeits, M. (2021).
\newblock Statistical postprocessing of wind speed forecasts using
  convolutional neural networks.
\newblock {\em Mon. Weather Rev.}, {\bf 149}(4):1141--1152.

\bibitem[Wilks, 2019]{wilks19}
Wilks, D.~S. (2019).
\newblock {\em Statistical Methods in the Atmospheric Sciences}.
\newblock Elsevier, Amsterdam, 4th edition.

\end{thebibliography}

\end{document}